\documentclass[onefignum,onetabnum]{siamonline190516}
\usepackage{amsmath}
\usepackage{amssymb}

\usepackage{lipsum}
\usepackage{amsfonts}
\usepackage{graphicx}
\usepackage{epstopdf}
\usepackage{algorithmic}
\ifpdf
  \DeclareGraphicsExtensions{.eps,.pdf,.png,.jpg}
\else
  \DeclareGraphicsExtensions{.eps}
\fi

\usepackage{enumitem}
\setlist[enumerate]{leftmargin=.5in}
\setlist[itemize]{leftmargin=.5in}

\newsiamremark{remark}{Remark}
\newsiamremark{hypothesis}{Hypothesis}
\crefname{hypothesis}{Hypothesis}{Hypotheses}
\newsiamthm{claim}{Claim}

\headers{Towards Convectons in the Supercritical Regime}{J. Tumelty, C. Beaume and A. M. Rucklidge}

\title{Towards Convectons in the Supercritical Regime: Homoclinic Snaking in Natural Doubly Diffusive Convection\thanks{Submitted to the editors DATE.
\funding{This work was supported by the Leeds\textendash{}York Natural Environment Research Council (NERC) Doctoral Training Partnership (DTP) SPHERES under grant NE/L002574/1 }}}

\author{J. Tumelty, C. Beaume and A. M. Rucklidge\thanks{Department of Applied Mathematics, University of Leeds, Leeds, LS2 9JT
		(\email{mmjt@leeds.ac.uk, C.M.L.Beaume@leeds.ac.uk, A.M.Rucklidge@leeds.ac.uk}).}}

\usepackage{amsopn}

\usepackage{cleveref}
\ifpdf
\hypersetup{
  pdftitle={Towards Supercritical Convectons: Homoclinic Snaking in Natural Doubly Diffusive},
  pdfauthor={J. Tumelty, C. Beaume, and A. M. Rucklidge}
}
\fi

\begin{document}

\maketitle

\begin{abstract}
Fluids subject to both thermal and compositional variations can undergo doubly diffusive convection when these properties both affect the fluid density and diffuse at
different rates.
A variety of patterns can arise from these buoyancy-driven flows, including spatially localised states known as convectons, which consist of convective fluid motion localised within a background of quiescent fluid.
We consider these states in a vertical
slot with the horizontal temperature and solutal gradients providing competing effects to the
fluid density while allowing the existence of a conduction state. 
In this configuration,
convectons have been studied with specific parameter values where the onset of convection is subcritical, and the states have been found to lie on
a pair of secondary branches that undergo homoclinic snaking in a parameter regime below the onset of linear
instability. 
In this paper, we show that convectons persist into parameter regimes in which the primary bifurcation is supercritical and there is no bistability, despite coexistence between the stable conduction state and large-amplitude convection.
We detail this transition by considering spatial dynamics and observe how the structure of the secondary branches becomes increasingly complex owing to the increased role of inertia at low Prandtl numbers.
\end{abstract}



\section{Introduction}
Fluids subject to both thermal and compositional variations can undergo doubly diffusive convection when both temperature and composition affect the fluid density and diffuse at different rates.
This situation is common in a variety of physical settings including oceanography \cite{huppert1981,schmitt1994double}, astrophysical flows \cite{garaud2018,garaud2021double,spiegel1969semiconvection}, crystallisation processes \cite{huppert1984,wilcox1993} or in the fluid dynamics that take place at the core-mantle boundary \cite{hansen1988numerical,lay2008}.
Doubly diffusive convection is noted for its ability to produce a variety of complex dynamics and patterns.
In particular, this system has often been used to investigate localisation in fluids \cite{batiste2006spatially,beaume2013convectons,beaume2018three,bergeon2008periodic,bergeon2008spatially,jacono2017localized,mercader2009convectons,mercader2011convectons,mercader2019,watanabe2012spontaneous}, where the spatially localised steady states typically take the form of convection rolls within a background conduction (motionless fluid) state and are referred to as convectons \cite{blanchflower1999magnetohydrodynamic}.

Localised states are most frequently studied in subcritical systems that display bistability between a trivial (conduction) state and a patterned (convection) state, with the Swift--Hohenberg equation, originally motivated by modelling thermal fluctuations in fluid convection \cite{swift1977}, considered the prototypical system \cite{burke2006localized,knobloch2015}.
In this system, localised states exist on branches that oscillate over a well-defined range of subcritical parameter values known as the pinning region.
This process is referred to as homoclinic snaking \cite{woods1999heteroclinic} and results in the coexistence of a multiplicity of localised states for the same parameter values.
The study of localised states in fluids has unsurprisingly revolved around these parameter values where subcriticality and bistability are observed \cite{batiste2006spatially, beaume2013convectonsrot,dawes2008localized,jacono2012spatially,mercader2009convectons, mercader2011convectons,richter2005two,schneider2010snakes,watanabe2012spontaneous} but these characteristics are not, in fact, necessary for spatial localisation.
Recent studies on systems involving a conserved quantity, including magnetoconvection \cite{cox2003instability,dawes2008localized,jacono2011magnetohydrodynamic,jacono2012spatially}, rotating convection \cite{beaume2013convectonsrot}, vibrating granular or fluid layers \cite{dawes2010localized,pradenas2017slanted}, optics \cite{firth2007homoclinic} and phase-field crystal models \cite{thiele2013localized}, have shown that localised states may lie on slanted snaking branches that can extend outside of the bistable region or may bifurcate from supercritical primary branches via modulational instabilities \cite{beaume2013convectonsrot,dawes2008localized}.
Such behaviour can arise when the conserved quantity generates a positive feedback mechanism enabling the localised states to be maintained at lower parameter values than where domain-filling states can be found.

The potential existence of localised states when the primary bifurcation is supercritical in the absence of a conserved quantity is less well-understood and likely differs across systems.
In the quadratic-cubic Swift\textendash{}Hohenberg equation, for example, exponential asymptotics has shown that the pinning region becomes exponentially thin during the approach towards the codimension-two point where the criticality of the bifurcation changes \cite{chapman2009exponential} and consequently localised states cease to exist when the primary bifurcation becomes supercritical.
In the cubic-quintic-septic Swift\textendash{}Hohenberg equation, however, Knobloch et al. \cite{knobloch2019defectlike} showed that homoclinic snaking and localised states can persist into regimes where the primary bifurcation is supercritical and bistability between the trivial state and a large-amplitude periodic state exists.
Further, the Swift\textendash{}Hohenberg equations considered in these studies are variational and it is therefore unclear how relevant any predictions based on this them are to the non-variational systems, including that of doubly diffusive convection.

In this paper, we consider the natural doubly diffusive convection in a vertically extended domain that is driven by horizontal gradients of temperature and solutal concentration.
Previous studies of localisation in this system \cite{beaume2013convectons,bergeon2008periodic, bergeon2008spatially,jacono2017localized} were restricted to parameter values that provided both subcriticality and bistability.
Here, we characterise how the homoclinic snaking that was found in the subcritical regime \cite{bergeon2008spatially} changes as the system is driven towards supercriticality by decreasing the Prandtl number \cite{beaume2022near}.
The key results we obtained are the discovery of convectons in the supercritical regime and the systematic characterisation of bifurcation diagrams during this transition.

The paper is organised as follows. 
In the next section, we detail the mathematical formulation of the system.
In section~3, we review the weakly nonlinear analysis of the system previously derived in \cite{beaume2022near} and extend the analysis into vertically extended domains, which provides insight into the origin of secondary branches of convectons. 
In section~4, we consider the nonlinear behaviour of the system and describe the successive stages that the snaking secondary branches undergo as the Prandtl number is decreased until the supercritical regime is reached.
We end, in section~5, with a discussion of the results.

\section{Mathematical formulation}\label{sec:mathematical_formulation}
\begin{figure}
	\centering
	\includegraphics{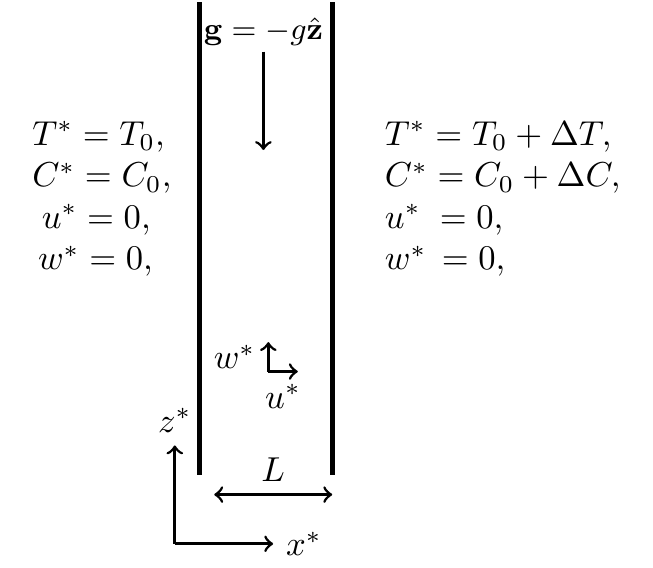}
	\caption{Sketch of the two-dimensional domain of interest together with the dimensional form of the boundary conditions.}
	\label{fig:domain}
\end{figure}

We consider natural doubly diffusive convection of an incompressible fluid in a two-dimensional domain with periodic boundary conditions imposed in the vertical direction.
The side walls are rigid, impermeable and maintained at fixed temperatures and solutal concentrations.
The left wall is held at temperature $T_0$ and solutal concentration $C_0$, while the right wall is held at a higher temperature ($T_0+\Delta T$, $\Delta T > 0$) and concentration ($C_0+\Delta C$, $\Delta C > 0$).
This configuration is depicted in Figure~\ref{fig:domain}. 

The system is governed by the Navier--Stokes equation for momentum, an incompressibility condition and advection-diffusion equations for both temperature and concentration.
Cross-diffusion due to the Soret and Dufour effects is not considered.
The imposed temperature and concentration differences are assumed to be sufficiently small so that the Boussinesq approximation can be applied, whereby density variations are neglected except in buoyancy terms.
We further assume that the fluid density depends linearly on both temperature and concentration:
\begin{equation}
\rho^* = \rho_0 + \rho_T(T^*-T_0) + \rho_C(C^* - C_0),
\end{equation}
where $\rho_0$ is the density of the fluid at temperature $T_0$ and concentration $C_0$ and $\rho_T<0$ (resp.~$\rho_C>0$) is the thermal (resp.~solutal) expansion coefficient.

We introduce the non-dimensional quantities:
\begin{equation}
\mathbf{x} = \frac{\mathbf{x}^*}{L},\quad t = \frac{t^*}{L^2/\kappa}, \quad \mathbf{u} = \frac{\mathbf{u}^*}{\kappa/L}, \quad T= \frac{T^*-T_0}{\Delta T}, \quad C = \frac{C^*-C_0}{\Delta C},\quad p = \frac{p^*}{\rho_0 \kappa \nu/L^2}, \label{eq:nondim}
\end{equation}
where $L$ is the wall separation, $\kappa$ is the rate of thermal diffusivity and $\nu$ is the kinematic viscosity.
The non-dimensional governing equations for the fluid velocity $\mathbf{u} = u \mathbf{\hat{x}} + w \mathbf{\hat{z}}$, the pressure $p$, the temperature $T$ and the concentration $C$ thus read:
\begin{align}
\dfrac{1}{Pr} \left(\dfrac{\partial \mathbf{u}}{\partial t} + \mathbf{u}\cdot \nabla \mathbf{u}\right) &= -\nabla p +\nabla^2 \mathbf{u} + Ra\left(T+NC\right)\mathbf{\hat{z}} , \label{eq:NS}\\
\nabla \cdot \mathbf{u} &= 0, \label{eq:incomp}\\
\dfrac{\partial T}{\partial t} + \mathbf{u}\cdot \nabla T &= \nabla^2 T, \label{eq:T}\\
\dfrac{\partial C}{\partial t} + \mathbf{u}\cdot \nabla C &= \frac{1}{Le}\nabla^2 C, \label{eq:C}
\end{align}
where $\mathbf{\hat{z}}$ is the vertical ascending unit vector and where we have introduced the following dimensionless parameters:
\begin{eqnarray}
\textrm{the Prandtl number~}&Pr &= \frac{\nu}{\kappa},\\
\textrm{the Rayleigh number~}&Ra &= \frac{gL^3|\rho_T|\Delta T}{\rho_0\nu \kappa},\\
\textrm{the buoyancy ratio~}&N &= \frac{\rho_C \Delta C}{\rho_T \Delta T} \quad\textrm{and}\\
\textrm{the Lewis number~}&Le &= \frac{\kappa}{D},
\end{eqnarray}
where $D$ is the rate of solutal diffusivity and $g$ is the standard acceleration due to gravity. 

We consider a domain of non-dimensional vertical extent $L_z = 12\lambda_c$, where $\lambda_c\approx 2.48$ is the wavelength of the critical eigenmode from the primary bifurcation of the conduction state for all values of the Prandtl and Lewis numbers \cite{beaume2022near,ghorayeb1997double,xin1998bifurcation}.
The non-dimensional boundary conditions read:
\begin{align}
u = 0,\quad w = 0, \quad \frac{\partial p}{\partial x} = \frac{\partial^2 u}{\partial x^2}, \quad T = 0, \quad C = 0& \quad \text{ on } \quad x = 0, \label{eq:bc0}\\
u = 0,\quad w = 0,\quad \frac{\partial p}{\partial x} = \frac{\partial^2 u}{\partial x^2},\quad T = 1, \quad C = 1& \quad \text{ on }\quad x = 1, \label{eq:bc1}
\end{align}
together with periodic boundary conditions in the $z$ direction and where the pressure boundary condition is the projection of the Navier--Stokes equation on the boundary.

Here, we restrict our attention to the case $N = -1$, where the full system (\ref{eq:NS}--\ref{eq:C}, \ref{eq:bc0}, \ref{eq:bc1}) admits the steady conduction state with linear temperature and concentration profiles between the side walls:
\begin{equation}
\mathbf{u}=\mathbf{0},\quad T = x,\quad C = x. \label{eq:conduction_state}
\end{equation}
Details of the system when ${N\neq-1}$, which leads to the conduction state being replaced by a base flow, are considered in \cite{tumelty2022localised}.

Numerical continuation of branches in the doubly diffusive convection system was carried out using a spectral element numerical method based on a Gauss--Lobatto--Legendre discretisation and supplemented by Stokes preconditioning \cite{beaume2017adaptive,bergeon2002, tuckerman2018}.
The domain was discretised using $24$ spectral elements with $25$ nodes in both the $x$ and $z$ directions.
The results are presented using bifurcation diagrams showing the kinetic energy:
\begin{equation}
E = \displaystyle\frac{1}{2} \int_0^1\int_0^{L_z}\left(u^2+w^2\right)\, dx\,dz,
\end{equation}
of steady states as a function of the Rayleigh number, which is treated as the bifurcation parameter.
Profiles of these steady states are depicted primarily using contour plots of the streamfunction $\psi$, where ${(u,w) = (-\psi_z,\psi_x)}$, while relationships between different steady states at the same parameter values are shown using spatial dynamics, where the vertical velocity of the state along the vertical line $x \approx 0.746$ is plotted against the horizontal velocity at the same point.

\section{Finding secondary bifurcations from the primary branch}\label{sec:5_secbif}
Previous studies of convectons in natural doubly diffusive convection with periodic boundary conditions focused on ${Le=11}$, ${Pr = 1}$ \cite{bergeon2008periodic, bergeon2008spatially} and found that convectons lie on a pair of secondary branches that bifurcate from a secondary bifurcation of the subcritical primary branch.
We start by deriving a general expression for the location of these small-amplitude secondary bifurcations, when they exist.

Our starting point for this analysis is the Ginzburg\textendash{}Landau equation derived by performing a weakly nonlinear analysis of the full nonlinear system (\ref{eq:NS}\textendash{}\ref{eq:bc1}) around the primary bifurcation of the conduction state \cite{beaume2022near}.
After introducing a small parameter ${\epsilon\ll1}$ used to define the slow temporal scale, ${T=\epsilon^2 t}$,  the long (vertical) spatial scale, ${Z=\epsilon z}$ and to quantify the deviation away from the primary bifurcation at $Ra_c$ via ${Ra = Ra_c + \epsilon^2r}$, the following equation is obtained:
\begin{equation}
A_{T} = a_1rA+a_2|A|^2A + a_3A_{ZZ}. \label{eq:5_GLE} 
\end{equation} 
This equation described the long spatial and slow temporal evolutions of the amplitude $A$ of the linear correction to the conduction state with wavenumber $k_c$.
The coefficients $a_1$ and $a_3$ are strictly positive provided that ${Le\neq1}$, whereas the sign of the coefficient $a_2$ depends upon the Prandtl and Lewis numbers and can either be positive, corresponding to a subcritical primary bifurcation, or negative, corresponding to a supercritical primary bifurcation.

There are three types of steady solutions of (\ref{eq:5_GLE}) that we are particularly interested in, namely:
\begin{equation}
A=0,
\end{equation}
which is valid for all $r$ and corresponds to the conduction state;
\begin{equation}
A = \left(-\frac{a_1r}{a_2}\right)^{1/2}e^{i\chi}, \label{eq:5_unifA}
\end{equation}
which is valid provided that ${a_1r/a_2 < 0}$ and corresponds to states of small-amplitude spatially periodic convection with arbitrary phase $\chi$; and
\begin{equation}
A = \left(-\frac{2a_1r}{a_2}\right)^{1/2}\text{sech}\left(\left(-\frac{a_1r}{a_3}\right)^{1/2}Z\right)e^{i\chi}, \label{eq:5_locA}
\end{equation}
which only exists in infinite domains and is valid provided that both $a_1r/a_2 < 0$ and\linebreak $a_1r/a_3 < 0$.
This solution corresponds to small-amplitude convection states with long spatial modulation.
While the phase $\chi$ appears to be arbitrary in (\ref{eq:5_locA}), including beyond-all-orders effects results in spatial locking between the two spatial scales $z$ and $Z$ and fixes the phase to either ${\chi =0}$ or ${\chi=\pi}$ in the Swift\textendash{}Hohenberg equation \cite{chapman2009exponential}, with the same likely being true in natural doubly diffusive convection \cite{bergeon2008periodic}. 
Similarly to the Swift\textendash{}Hohenberg equation \cite{burke2006localized}, the two types of modulated states extend towards lower $r$ and develop into fully localised states.

\begin{figure}
	\centering
	\includegraphics[trim={0 4.25cm 0 0},clip]{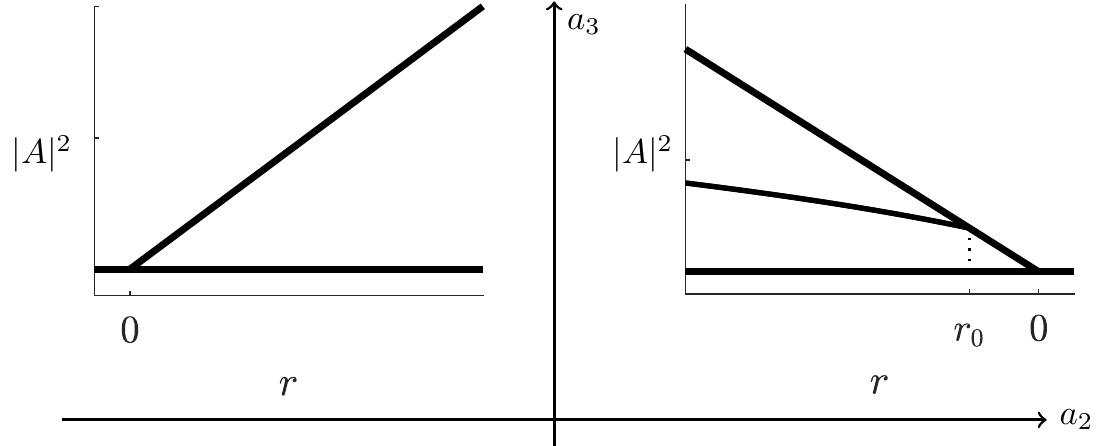}
	\caption[Sketches of the nature of the primary and secondary bifurcations of the Ginzburg\textendash{Landau} equation in different regimes of $(a_2,a_3)$ parameter space]{Sketches of the nature of the primary and secondary bifurcations of the Ginzburg\textendash{Landau} equation (\ref{eq:5_GLE}) for ${a_2>0}$ and ${a_2<0}$ when both ${a_1> 0}$ and ${a_3>0}$. 
		The secondary bifurcation occurs at ${r = r_0}$ and the direction of branching is obtained numerically. }
	\label{fig:5secbifa}
\end{figure}

In finite domains, however, the latter states do not exist and the two branches of localised states are instead found to originate from an Eckhaus instability of the primary branch \cite{bergeon2008eckhaus}.		
By performing a linear stability analysis on the non-trivial, uniform solutions to the Ginzburg\textendash{}Landau equation (\ref{eq:5_unifA}), we can determine when this secondary bifurcation occurs. 
We first move into the frame of reference of the non-trivial periodic state with ${\chi = 0}$ and make a small, real perturbation $b$ that is even in $Z$ both for simplicity and for connection with equation (\ref{eq:5_locA}):
\begin{equation}
A(Z,T) = \left( -\dfrac{a_1r}{a_2}\right)^{1/2} + b(Z,T).
\end{equation}
In this frame of reference, the Ginzburg\textendash{}Landau equation becomes
\begin{equation}
b_{T} = -2a_1rb + 3\left(-a_1a_2r\right)^{1/2}b^2 + a_2b^3 + a_3b_{ZZ}.
\end{equation}
Linearising this equation about a potential secondary bifurcation at $r=r_0$, we find that the leading-order deviation from the constant amplitude state satisfies
\begin{equation}
b_{T} = -2a_1r_0b + a_3 b_{ZZ} + \mathcal{O}(\epsilon^2),
\end{equation}
which has solution:
\begin{equation}
b = \epsilon B_1e^{\lambda T}\cos(lZ) + \mathcal{O}(\epsilon^2), 
\end{equation}
where $B_1$ is the amplitude of the perturbation and where the wavenumber $l$, growth rate $\lambda$ and the location of a secondary bifurcation $r_0$ are related via
\begin{equation}
r_0 \approx -\frac{\lambda + l^2a_3}{2a_1}. \label{eq:5_locsecondarybifgen}
\end{equation}
Since $l$ is real, relation (\ref{eq:5_locsecondarybifgen}) requires that ${a_1r_0/a_3 < 0}$ at a stationary bifurcation (${\lambda = 0}$).
Combining this condition with ${a_1r_0/a_2 < 0}$, which is necessary for the existence of a uniform, steady, finite-amplitude solution $A$, we find that such a small-amplitude secondary bifurcation can only occur when $a_2$ and $a_3$ take the same sign.
Thus, since ${a_3>0}$ for all values of the Prandtl and Lewis numbers \cite{beaume2022near}, this linear stability theory predicts that the primary branch only undergoes a small-amplitude secondary bifurcation when this branch is subcritical, as illustrated in figure~\ref{fig:5secbifa}. 

To determine the location of the first of the secondary bifurcations, we note that the periodic boundary conditions in the vertical direction discretises the possible wavenumbers to ${l =2\pi n/(\epsilon L_z)}$ for $n = 1,2, ...$, which means that the first of these secondary bifurcations occurs when ${l = 2\pi/(\epsilon L_z)}$. 
From (\ref{eq:5_locsecondarybifgen}), this wavenumber gives the expected location of the first secondary bifurcation as
\begin{equation}
Ra \approx Ra_c-\frac{2\pi^2}{L_z^2}\frac{a_3}{a_1}. \label{eq:5_locsecondarybif}
\end{equation}
Using the parameter dependence of the coefficients $a_1$ and $a_3$ from \cite{beaume2022near}, this location may be expressed more generally as
\begin{equation}
Ra \approx \frac{1}{|1-Le|}\left(6509-\frac{2\pi^2}{L_z^2}\frac{\delta}{|\gamma_1|}\right), \label{eq:5_locsecondarybif_pardep}
\end{equation}
where $\delta$ and $\gamma_1$ are constants defined in \cite{beaume2022near}, which are independent of both the Prandtl and Lewis numbers.

\begin{figure}
	\centering
	\includegraphics{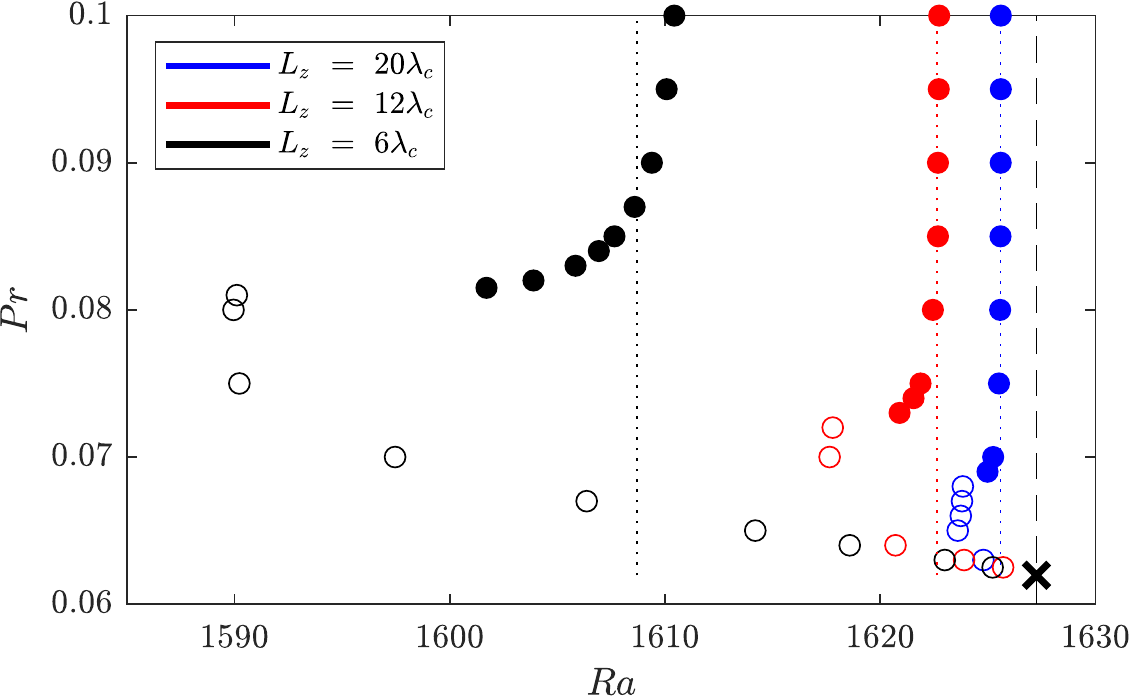}
	\caption[Location of secondary bifurcation for different values of the Prandtl number when ${Le = 5}$ in domains with ${L_z = 20\lambda_c}$, ${L_z = 12\lambda_c}$ and ${L_z = 6\lambda_c}$]
	{Location of the first secondary bifurcation of the primary branch for different values of the Prandtl number when ${Le = 5}$ in domains with ${L_z = 20\lambda_c}$ (blue), ${L_z = 12\lambda_c}$ (red) and ${L_z = 6\lambda_c}$ (black). 
		Stationary (Hopf) bifurcations are depicted using filled (open) circles.
		The vertical dotted lines indicate the theoretical location of the secondary bifurcation (\ref{eq:5_locsecondarybif_pardep}) obtained via linear stability analysis.
		The vertical dashed line at ${Ra\approx 1627.26}$ denotes the location of the primary bifurcation and the black cross on this line at ${Pr = Pr_c\approx 0.062}$ indicates the Prandtl number for the codimension-two point, below which the primary bifurcation is supercritical.}
	\label{fig:5_secbifloclz}
\end{figure}

The above analysis holds for sufficiently large domains away from the codimension-two point $ (Ra_c,Pr_c)$, where the primary bifurcation changes criticality. 
Figure~\ref{fig:5_secbifloclz} confirms this by comparing the theoretical location of the first secondary bifurcation (\ref{eq:5_locsecondarybif_pardep}) with that obtained numerically for a range of domain sizes and Prandtl numbers. 
In the domains with ${L_z = 12\lambda_c}$ and ${L_z = 20\lambda_c}$, where ${\lambda_c=2\pi/k_c}$ is the critical wavelength, the stationary bifurcations approach the theoretical limit (dotted) as the Prandtl number increases.
However, decreasing the Prandtl number towards $Pr_c$ in each domain, the stationary secondary bifurcations move away from this theoretical limit and towards lower Rayleigh numbers. 
This bifurcation proceeds to collide with a second stationary bifurcation and becomes an oscillatory bifurcation before moving towards the primary bifurcation as the codimension-two point is approached.

To capture the additional details in finite domains near the codimension-two point seen in figure~\ref{fig:5_secbifloclz}, an alternative rescaling that leads to a higher order Ginzburg\textendash{}Landau equation should be used \cite{dawes2009modulated,kao2012weakly}.
We have not considered such an analysis here because the focus of this work is on the large-amplitude snaking behaviour.
Nevertheless, the results obtained via numerical continuation seen in section~4 indicate that a supercritical primary branch does not undergo a small-amplitude secondary bifurcation.

\section{From subcritical to supercritical convectons}\label{sec:5_trans}
One of the main results obtained in \cite{beaume2022near} was the existence of parameter regimes where the primary bifurcation is supercritical and the stable conduction state coexists with states of spatially periodic convection.
This coexistence arises as the by-product of the evolution of the primary branch as it transitions from being subcritical at high Prandtl numbers to supercritical at low Prandtl numbers and undergoes a cusp bifurcation, where two new saddle nodes emerge.
One of the three saddle nodes terminates at the codimension-two point $(Ra_c,Pr_c)$, resulting in an `S'-shaped primary branch below the critical Prandtl number where the branch initially heads towards supercritical Rayleigh numbers, turns around at a saddle node and heads back towards lower Rayleigh numbers, before turning around at a second saddle node and proceeding to large Rayleigh numbers and amplitude.
The second saddle node lies in the subcritical region ${Ra < Ra_c}$ for certain parameter values, which leads to the coexistence between the stable conduction state and states of spatially periodic convection.
Despite the primary branch not undergoing a small-amplitude secondary bifurcation in the supercritical regime, we find that supercritical systems of natural doubly diffusive can admit convectons that are supported by this coexistence.
In the remainder of this paper, we are not only interested in details of these convectons in the supercritical regime, but also in understanding their origin in relation to how the typical snaking branches that exist in the subcritical regime develop as the Prandtl number decreases and the primary bifurcation becomes supercritical.

\begin{figure}
	\centering
	\includegraphics[width=0.99\linewidth]{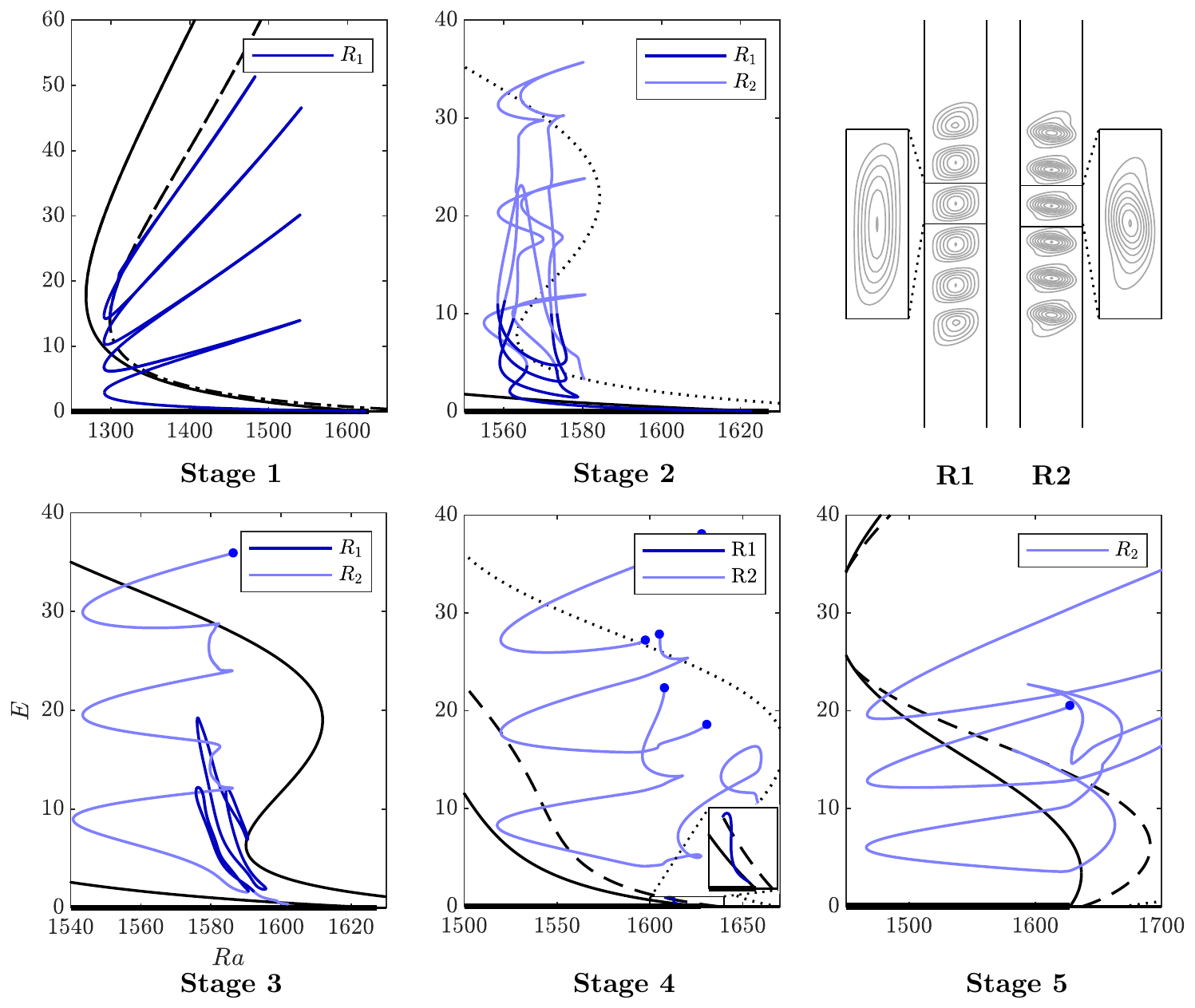}
	\caption[Summary of the five different stages that are seen in the transition from the primary bifurcation being subcritical to supercritical]{Summary of the five different stages that are seen in the transition from the primary bifurcation being subcritical to supercritical. 
		Illustrative bifurcation diagrams for the five stages: Stage~1 (${Pr = 1}$); Stage~2 (${Pr = 0.11}$); Stage~3 (${Pr = 0.102}$); Stage~4 (${Pr = 0.09}$); and Stage~5 (${Pr = 0.06}$).
		Branches of convectons with an even number of rolls are shown in blue and the branch segments predominantly corresponding to the two types of rolls are separated by showing convectons with R1 rolls in dark blue and R2 rolls in light blue.
		The branches have been terminated at the blue dots.
		The primary branches PN, consisting of steady spatially periodic states with N rolls, are shown in black with the following line styles: P12 (solid), P11 (dashed), P10 (dotted) and P9 (dash-dotted).
		Streamfunctions of convectons distinguishing the two types of roll (R1 and R2) are shown in the top right, with the smaller panels showing streamfunctions for a single convection roll of each type, which have been magnified so that the rolls are to-scale.
		Further details of the distinction between rolls of type R1 and R2 are provided in figures~\ref{fig:5_vortPr1} and \ref{fig:5_vortPr012}.
		}
	\label{fig:5_summary}
\end{figure}

We recall that we focus on the steady state dynamics in a domain that is periodic in the vertical direction with a period $L_z \approx 12\lambda_c$, where $\lambda_c\approx 2.48$ is the wavelength of the critical eigenmode from the primary bifurcation of the conduction state for all values of the Prandtl and Lewis numbers \cite{ghorayeb1997double,xin1998bifurcation}.
We further fix ${Le=5}$, where the primary bifurcation at ${Ra_c \approx 1627.26}$ changes criticality at ${Pr_c \approx 0.062}$.
These parameters are chosen so that the system admits convectons and so that both the sub- and supercritical regimes are numerically accessible.
This allows us to explore how the structure of the snaking branches of convectons changes as the Prandtl number varies between ${Pr = 1}$ and ${Pr = 0.06}$ and the primary bifurcation changes from being subcritical to supercritical.
We characterise this transition in five stages depicted using representative bifurcation diagrams in figure~\ref{fig:5_summary} and briefly summarised in the following paragraph. 

The first of these stages is the subcritical regime for moderate and large Prandtl numbers (${Pr \gtrsim 0.15}$), subsequently referred to as Stage~1, where we find that the convectons lie on a pair of branches that undergo homoclinic snaking, similar to that found by Bergeon and Knobloch \cite{bergeon2008spatially}.
The structure of the snaking branches becomes increasingly complex during Stage~2 (${0.11 \lesssim Pr\lesssim 0.15}$) as rolls in convectons change between buoyancy-driven rolls, hereafter referred to as R1, like those in Stage~1, and rolls driven by a balance between buoyancy and inertia, hereafter referred to as R2.
Within Stage~3 (${0.102 \lesssim Pr\lesssim 0.11}$), each snaking branch breaks up into a main branch containing convectons with R2 rolls and a collection of isolas containing convectons with R1 rolls.
During Stage~4 (${Pr_c \approx 0.062 \lesssim Pr\lesssim 0.102}$), the main branch further breaks up into a set of disconnected branch segments, while the set of isolas connect together to give small-amplitude snaking between a pair primary branches.
When the primary bifurcation is supercritical in Stage~5 (${Pr\lesssim Pr_c\approx 0.062}$), this small-amplitude snaking no longer exists, but the large-amplitude snaking persists.

		\subsection{Stage 1: Typical homoclinic snaking}
\begin{figure}
	\centering	
	\includegraphics[width=0.99\linewidth]{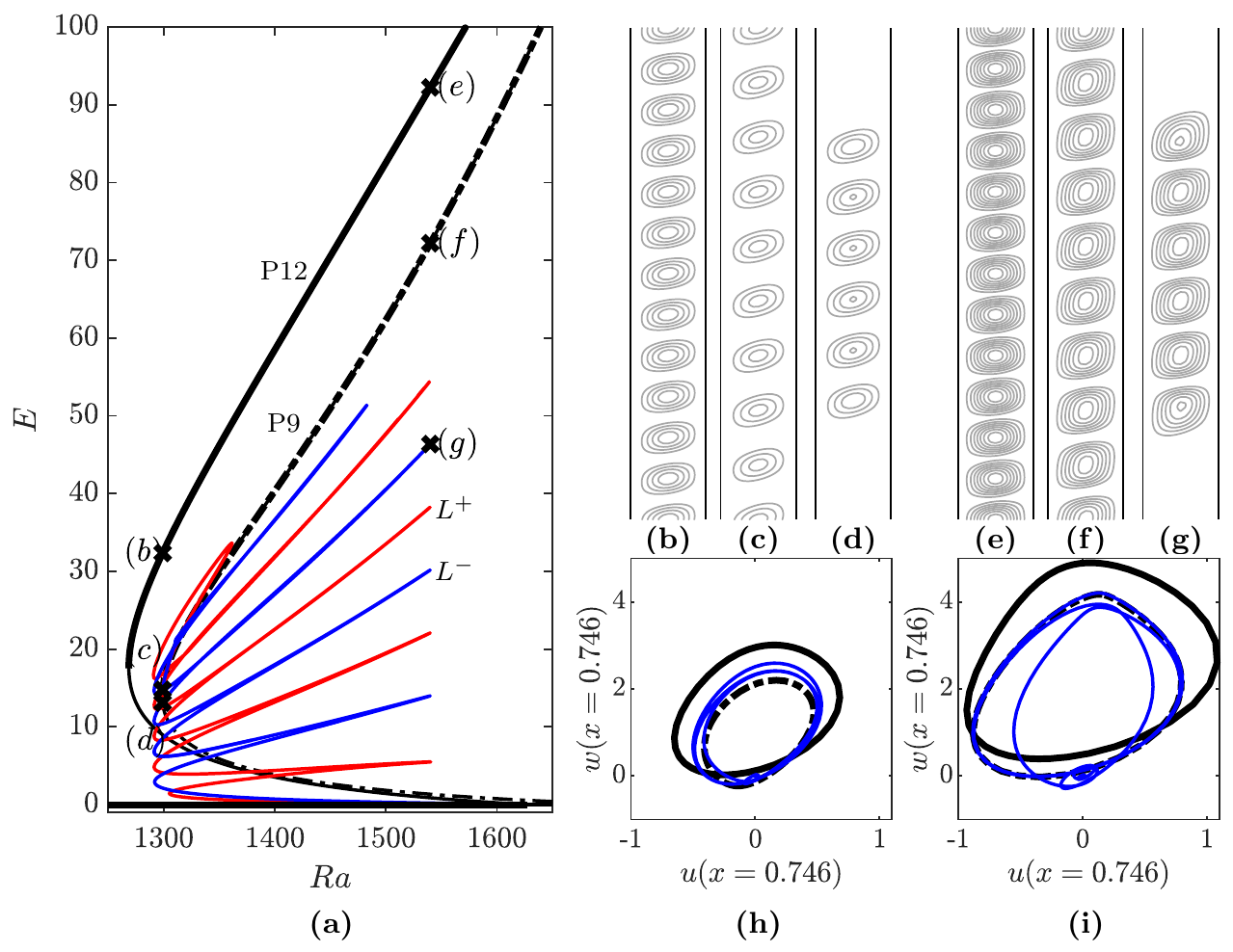}
	\caption[Homoclinic snaking and steady states for ${Pr = 1}$ ]{Homoclinic snaking and steady states for ${Pr = 1}$ (Stage~1). 
		(a) Bifurcation diagram showing P12 (black solid), P9 (black dash-dotted), $L^+$ (red solid) and $L^-$ (blue solid). 
		Thick (thin) lines indicate stable (unstable) spatially periodic states on the primary branches. The states marked with a cross are represented in the other panels.
		(b)\textendash{}(d) Streamfunctions of states at ${Ra = 1299}$ on (b) P12, (c) P9, (d) $L^-$ and (e)\textendash{}(g) states at ${Ra = 1540}$ on (e) P12, (f) P9 and (g) $L^-$. 
		Phase-space representation of the marked states at (h) ${Ra = 1299}$ and (i) ${Ra =1540}$, where (b) and (e) are shown via the black solid lines, (c) and (f) are shown via the black dash-dotted lines and (d) and (g) are shown via the blue solid lines.}
	\label{fig:5_Pr1}
\end{figure}
\begin{figure}
	\centering	
	\includegraphics[width=0.99\linewidth]{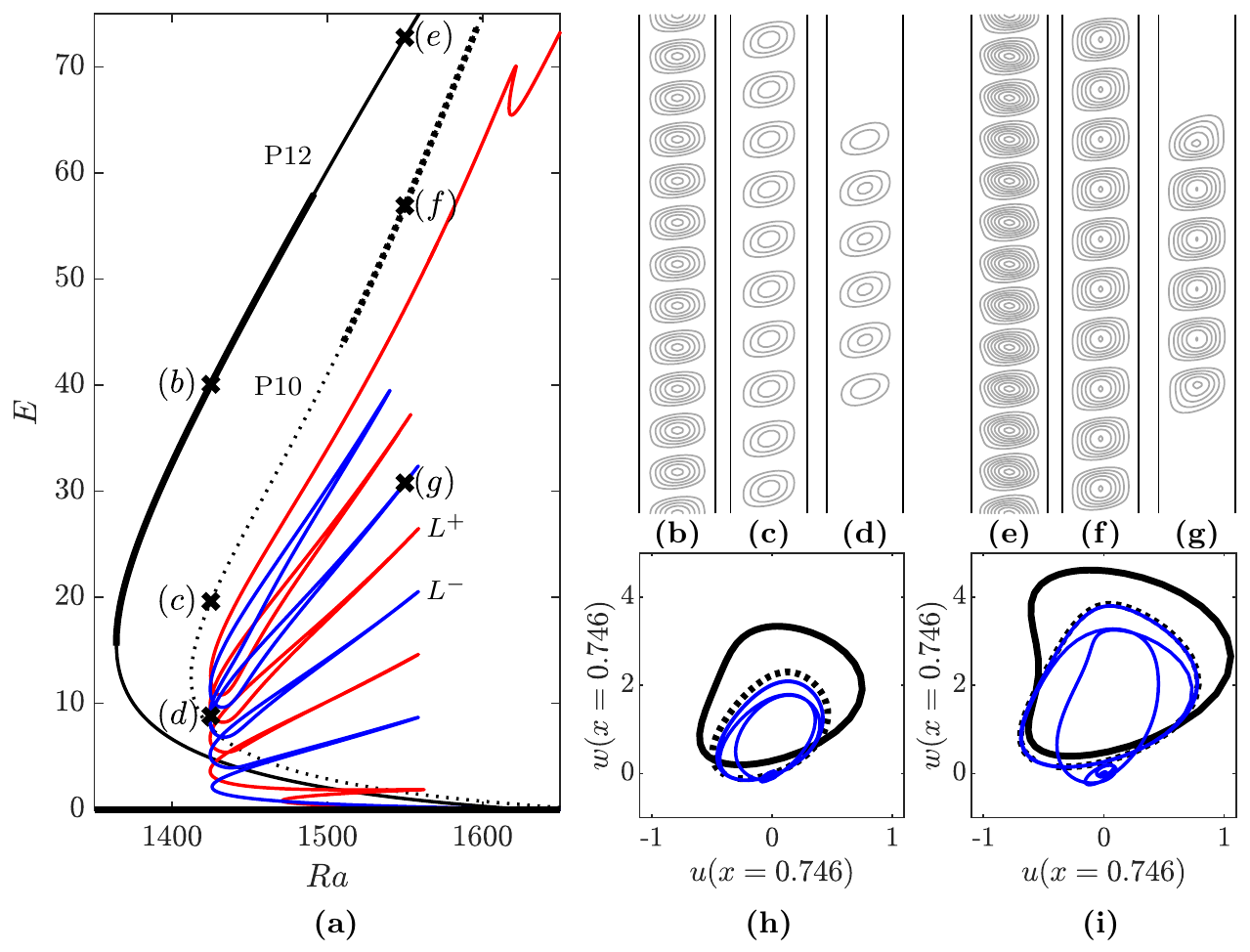}
	\caption[Homoclinic snaking and steady states for ${Pr = 0.2}$]{Homoclinic snaking and steady states for ${Pr = 0.2}$ (Stage~1). 
		(a) Bifurcation diagram showing P12 (black solid), P10 (black dotted), $L^+$ (red solid) and $L^-$ (blue solid). 
	  Thick (thin) lines indicate stable (unstable) spatially periodic states on the primary branches.
          The states marked with a cross are represented in the other panels.
		(b)\textendash{}(d) Streamfunctions of states at ${Ra = 1425}$ on (b) P12, (c) P10, (d) $L^-$ and (e)\textendash{}(g) states at ${Ra = 1550}$ on (e) P12, (f) P10 and (g) $L^-$. 
		Phase-space representation of the marked states at (h) $R{a = 1425}$ and (i) ${Ra =1550}$, where (b) and (e) are shown via the black solid lines, (c) and (f) are shown via the black dotted lines and (d) and (g) are shown via the blue solid lines.}
	\label{fig:5_Pr02}
\end{figure}

We start by describing the large Prandtl number behaviour found within Stage 1 (${Pr > 0.15}$), which is illustrated in figures~\ref{fig:5_Pr1} and \ref{fig:5_Pr02}.
Here, we recover the homoclinic snaking similar to that found in the Swift\textendash{}Hohenberg equation (cf. \cite{burke2006localized}), which was first identified in this system in \cite{bergeon2008spatially}. 
For these large Prandtl numbers, the stable conduction state first destabilises subcritically at ${Ra_c \approx 1627}$ to an eigenmode with twelve pairs of counterrotating rolls. 
The resulting primary branch, named P$12$, extends towards lower Rayleigh numbers before regaining stability at a subcritical saddle node (${Ra\approx 1269}$ for ${Pr = 1}$ in figure~\ref{fig:5_Pr1}(a) or ${Ra\approx 1364}$ for ${Pr = 0.2}$ in figure~\ref{fig:5_Pr02}(a)).
The stable upper branch (shown in bold in figures~\ref{fig:5_Pr1}(a) and \ref{fig:5_Pr02}(a)) heads towards large Rayleigh numbers and may destabilise in a drift-pitchfork bifurcation, like that seen at ${Ra\approx 1491}$ for ${Pr = 0.2}$ (figure~\ref{fig:5_Pr02}(a)). 

The conduction state later destabilises to eigenmodes with ${N\neq 12}$ pairs of counterrotating rolls, where the primary branches P$N$ ($N=11,10,9,\dots$) bifurcate.
These bifurcations may be either sub- or supercritical, depending upon the number of rolls and the Prandtl number.
The subsequent structure and stability of these branches are highly variable and will not be considered in detail here.
However, we should note that following each of the primary branches towards larger amplitude, the anticlockwise rolls within the steady states strengthen at the expense of the clockwise rolls.
Far from the primary bifurcation, the states on P$N$ thus consist of $N$ corotating, anticlockwise rolls (see figures~\ref{fig:5_Pr1}(b), (e) and \ref{fig:5_Pr02}(b), (e) for ${N = 12}$, figures~\ref{fig:5_Pr1}(c), (f) for ${N = 9}$ and figures~\ref{fig:5_Pr02}(c), (f) for ${N = 10}$).

Shortly after the primary bifurcation at $Ra_c$, two secondary branches, $L^-$ and $L^+$, bifurcate subcritically from a modulational instability of P12 at $Ra \approx 1622.8$, which is in agreement with the theoretical result (\ref{eq:5_locsecondarybif_pardep}).
As these secondary branches head towards lower Rayleigh numbers, the steady states that lie on them become increasingly modulated in space while exhibiting the same roll growth disparity as the primary branch states. 
This leads to states on $L^+$ consisting of a single isolated anticlockwise roll and those on $L^-$ consisting of two isolated anticlockwise rolls by their respective first left saddle node. 
Both secondary branches proceed by undergoing homoclinic snaking over a finite range of Rayleigh numbers:
the interior rolls first strengthen as the branches go from left-to-right saddle nodes; 
a roll nucleates on either side of the existing convecton rolls at the right saddle nodes; 
and finally the outer rolls strengthen while the interior rolls weaken as the branches go from right-to-left saddle nodes.
The number of rolls in convectons therefore increases as the branches are followed towards larger amplitude until rolls nearly fill the domain.
When this point is reached, the secondary branches $L^+$ and $L^-$ either turn over and connect to a primary branch of periodic states (e.g., P9 when ${Pr = 1}$ and P10 when ${Pr = 0.2}$), or extend towards higher Rayleigh numbers, depending upon the preferred wavelength of the localised state \cite{bergeon2008eckhaus}.

\begin{figure}
	\centering	
	\includegraphics[width=0.99\linewidth]{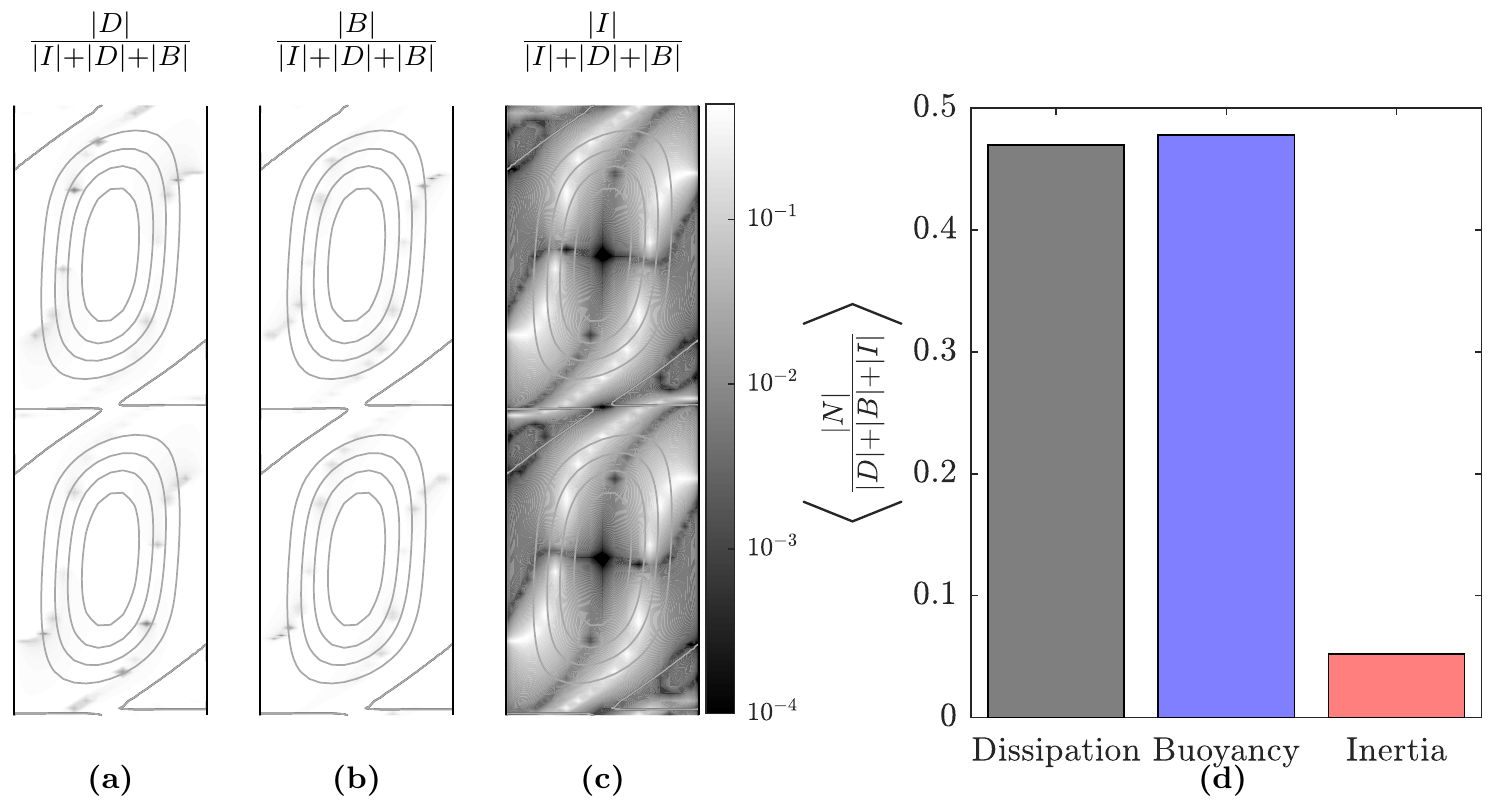}
	\caption[Vorticity balance for the two central rolls of the six-roll convecton with ${Pr = 1}$ and ${Ra = 1540}$]{Vorticity balance (\ref{eq:5_vortbalance}) for the two central rolls of the six-roll convecton  with ${Pr = 1}$ and ${Ra = 1540}$ (figure~\ref{fig:5_Pr1}(g)), which contains R1 rolls.
		Spatial dependence of the fraction that (a) the viscous dissipation term $|D|$, (b) the buoyancy term $|B|$ and (c) the inertial term $|I|$ contribute to the sum of terms ${|I| + |D| + |B|}$ at each point.
		The colour bar on the right indicates the logarithmic scale used: white indicates a dominant term, while grey and black indicate a subdominant term.
		Streamlines of the steady flow are superposed onto each plot.
		(d) Average fraction of each term in (a)\textendash{}(c) over the domain shown. }
	\label{fig:5_vortPr1}
\end{figure}

The central convecton rolls closely resemble those of individual rolls in a state at the same Rayleigh number on the upper branch segment of one of the primary branches, especially near the right saddle nodes (e.g., compare figures~\ref{fig:5_Pr1}(f) and (g) and figures~\ref{fig:5_Pr02}(f) and (g)).
The flow near the centre of these rolls is approximately a vertical shear flow that is nearly parallel to the side walls, whereas the flow is inclined upwards (resp.~downwards) towards the hotter (resp.~colder) wall at the bottom (resp.~top) of the roll.
These rolls are subsequently referred to as R1 and may be characterised by the inertial term (I) playing a subdominant role in the steady-state balance of the vorticity equation:
\begin{equation}
\underbrace{\vphantom{\left(\frac{\partial \Theta}{\partial x}\right)}\frac{1}{Pr} \left(\mathbf{u}\cdot \nabla\omega\right)}_{I}
- \underbrace{\vphantom{\left(\frac{\partial \Theta}{\partial x}\right)}\nabla^2 \omega}_{D} 
+ \underbrace{Ra \left(\frac{\partial T}{\partial x}- \frac{\partial C}{\partial x}\right)}_{B} 
= 0,\label{eq:5_vortbalance}
\end{equation}
where ${\omega= \hat{\mathbf{y}} \cdot\nabla \times \mathbf{u}}$. 
This is evidenced in figure~\ref{fig:5_vortPr1}, which shows the spatial dependence of the fraction that each term of the vorticity equation contributes compared to the sum of the absolute values of all terms at the same point, for the central two rolls of the six-roll convecton at ${Ra = 1540}$ and ${Pr = 1}$, shown in figure~\ref{fig:5_Pr1}(g). 
Panels (a) and (b) illustrate that viscous dissipation, $D$, and buoyancy, $B$, provide the balance in (\ref{eq:5_vortbalance}), while inertia, $I$, is subdominant, except in a a thin `S'-shaped white strip on each side of each roll.
This is further observed in panel (d), where the domain averaged fractions for both viscous dissipation and buoyancy are $0.47$ and $0.478$, respectively, while that for inertia is $0.05$.
Thus, at large Prandtl numbers, the anticlockwise convection rolls arise primarily via a balance between viscous dissipation and the buoyancy forcing.

The similarity between rolls in convectons and periodic states can also be illustrated by plotting different steady states at the same parameter values together in the phase space defined by the velocity components of the state along the vertical line ${x \approx 0.746}$.
In this phase-space representation, the conduction state becomes a single point located at the origin, whereas the spatially periodic states on primary branches become closed periodic orbits.
Selected states on the upper primary branches are shown by the black curves in figures~\ref{fig:5_Pr1}(h), (i) and \ref{fig:5_Pr02}(h), (i), where the line-style corresponds to the associated primary branch, i.e., black solid, dotted and dash-dotted orbits represent states on P12, P10 and P9, respectively.
The spatial trajectories of the six-roll convectons (figures~\ref{fig:5_Pr1}(d), (g), \ref{fig:5_Pr02}(d), (g)) are represented by the blue curves in these phase-space plots (figures~\ref{fig:5_Pr1}(h), (i), \ref{fig:5_Pr02}(h), (i)).
Each of these trajectories starts near the origin, as the flow is almost motionless away from the convecton rolls, and proceeds to spiral outwards as the weak rolls that constitute the front connecting the conduction state to convection rolls are traversed.
The trajectory then approaches a periodic orbit corresponding to a state on the upper branch of P9 (black dash-dotted) when ${Pr = 1}$ in figures~\ref{fig:5_Pr1}(h), (i) or P10 (black dotted) when $Pr = 0.2$ in figures~\ref{fig:5_Pr02}(h), (i) and follows this periodic orbit around four times as the four interior convecton rolls are traversed. 
The trajectory finally spirals back towards the origin as the front connecting the convection rolls to the conduction state is traversed.

Convecton rolls continue to have the form of R1 rolls and appear in phase space as trajectories that follow the periodic orbit for a state on the upper branch of P10 as the Prandtl number decreases through the remainder of Stage 1 (${0.15\lesssim Pr < 0.2}$).
The structure of the snaking branches prior to termination or large-amplitude behaviour also remains qualitatively unchanged over this interval.
\subsection{Stage 2: Emergence of inertia-dominated convectons}

\begin{figure}
		\centering
		\includegraphics[width=0.99\linewidth]{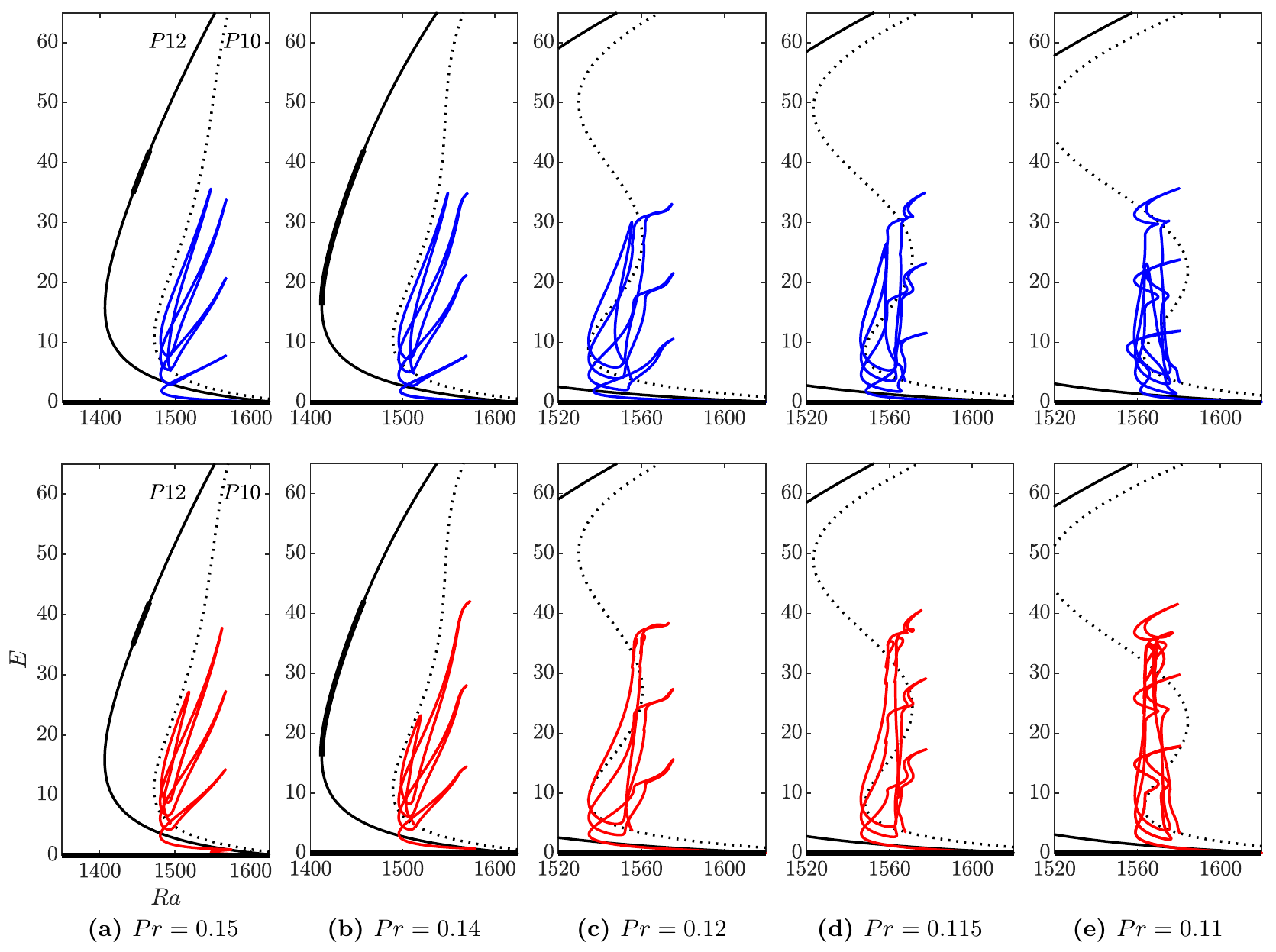}	
		\caption[Bifurcation diagrams showing how the structure of the secondary branches changes over Stage 2]{Bifurcation diagrams for (a) ${Pr= 0.15}$, (b) ${Pr=0.14}$, (c) ${Pr=0.12}$. (d) ${Pr=0.115}$ and (e) ${Pr=0.11}$, which show how the structure of the secondary branches changes over Stage 2.
			The marked branches include: the primary branches P12 (black solid) and P10 (black dotted) and secondary branches $L^-$ (blue branches in top row) and $L^+$ (red branches in bottom row).
			Stable (unstable) segments of the primary branches are indicated using bold (thin) lines.}
		\label{fig:5_bd2lz12le5pr011015}
	\end{figure}

The structure of the snaking branches changes across Stage 2 (${Pr\approx 0.15}$ to ${Pr \approx 0.11}$), as P10 undergoes a cusp bifurcation on its upper branch segment at ${Pr \approx 0.14}$ and ${Ra \approx 1547}$ (second column of figure~\ref{fig:5_bd2lz12le5pr011015}). 
This bifurcation introduces a pair of additional saddle nodes and branch segments on the P10 branch, as can be seen in the final three columns of figure~\ref{fig:5_bd2lz12le5pr011015}. 
Of particular interest are the two branch segments with positive gradient, which, in order of increasing kinetic energy, will hereafter be referred to as the second and upper branch segments of P10.
We shall see that the change in snaking structure arises because the form of the convecton rolls changes from R1, where inertia is subdominant, to a second form, hereafter referred to as R2, where inertia is part of the dominant balance.
This transition impacts the spatial dynamics in the following way: the homoclinic orbits associated with convectons transition from approaching periodic orbits corresponding to states on the second branch segment at low Rayleigh numbers to those on the upper branch segment at large Rayleigh numbers.
This change occurs despite both the second and upper branch segments being unstable to modulational and drift instabilities, respectively.
	\begin{figure}
		\centering
		\includegraphics[width=0.99\linewidth]{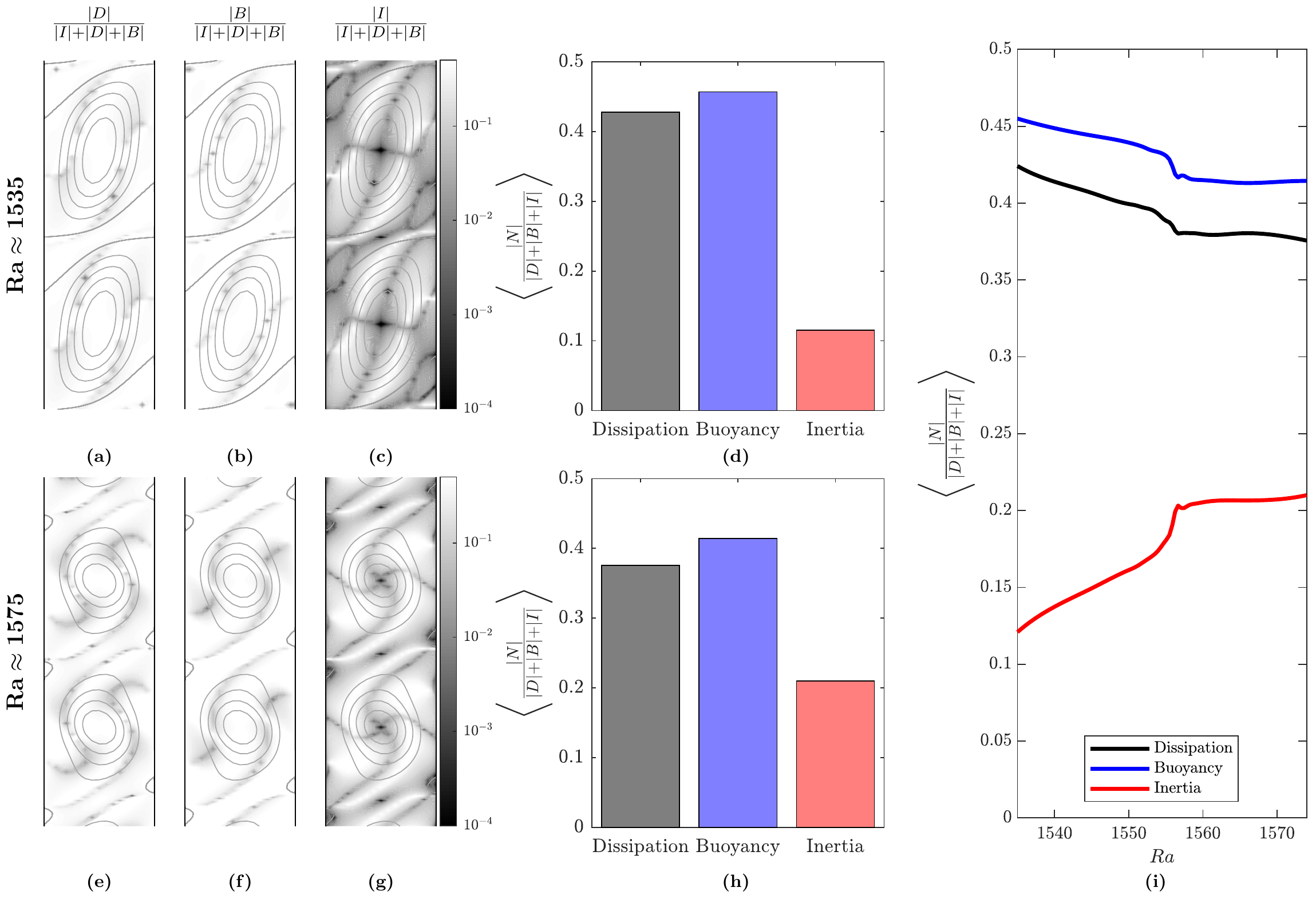}
		\caption[Vorticity balance for the central two-rolls of the six-roll convectons with ${Pr = 0.12}$ at the saddle nodes at ${Ra = 1535}$ and ${Ra = 1575}$ and along the connecting branch segment]{(a)\textendash{}(h) Similar to figure~\ref{fig:5_vortPr1}, the states are the six-roll convectons for ${Pr = 0.12}$ at: (a)\textendash{}(d) the left saddle node at ${Ra \approx 1535}$ (figure~\ref{fig:5_Pr012}(g)) with R1 rolls, and (e)\textendash{}(h) the right saddle node at ${Ra \approx 1575}$ (figure~\ref{fig:5_Pr012}(j)), with R2 rolls.
			(i) Domain averaged fraction of maximum term in the vorticity equation for each of viscous dissipation (black), buoyancy (blue) and inertial (red) terms over the branch segment between these left and right saddle nodes.
		}
		\label{fig:5_vortPr012}
	\end{figure}

The differences between the two types of rolls can be understood by considering how the vorticity balance (\ref{eq:5_vortbalance}) of steady convectons changes when the Rayleigh and Prandtl numbers are varied.
Figure~\ref{fig:5_vortPr012} presents such a comparison when the central two rolls of a six-roll convecton with ${Pr = 0.12}$ are considered between the left saddle node at ${Ra \approx 1535}$ (panels (a)\textendash{}(d)), where rolls are R1, and the right saddle node at ${Ra \approx 1575}$ (panels (e)\textendash{}(h)), where rolls are R2.
At the left saddle node, the spatial dependence of terms in the vorticity equation (figures~\ref{fig:5_vortPr012}(a)\textendash{}(c)) resembles that seen in figure~\ref{fig:5_vortPr1}.
Dissipation and buoyancy are dominant and respectively contribute approximately $42\%$ and $46\%$ of the domain-averaged values (figure~\ref{fig:5_vortPr012}(d)), while the inertial term is subdominant and contributes approximately $12\%$ of the balance.
The slight increase in inertial contributions between figures~\ref{fig:5_vortPr1}(c) and \ref{fig:5_vortPr012}(c) is seen through the widening of the white strips in the top left and bottom right regions of each roll and the overall decrease in intensity of the greyscale.
The inertial contribution increases further with increasing Rayleigh number along the branch segment corresponding to six-roll convectons when ${Pr = 0.12}$, as can be seen in figure~\ref{fig:5_vortPr012}(i).
This increase has a near uniform rate until ${Ra \approx 1555}$, as rolls strengthen whilst maintaining the form of R1 rolls.
At ${Ra \approx 1556}$, the fraction that inertia (buoyancy, viscous dissipation) contributes to the balance rapidly increases (decreases) to 0.20 (0.42, 0.38) and remains approximately at this level towards the right saddle node at ${Ra \approx 1575}$.
The rapid change, which we will explore further in the following paragraph, is associated with the central rolls in the convecton changing from R1 to R2, where inertia is part of the dominant balance.

Aside from the different proportions in the vorticity balance, the two types of rolls also have qualitatively different structures, as evidenced by the streamfunctions representations in figure~\ref{fig:5_vortPr012}.
We can see that R2 rolls (figures~\ref{fig:5_vortPr012}(e)\textendash{}(g)) are smaller yet stronger than R1 rolls (figures~\ref{fig:5_vortPr012}(a)\textendash{}(c)) and also that the long axis of the near elliptical streamlines at the centre of the roll have different orientations.
In R1 rolls, this axis inclines upwards towards the hotter right side wall, while the long axis at the centre of R2 rolls inclines upwards towards the colder left side wall.
The latter observation is seen most clearly using the grey lines that originate from the centre of each roll in panels (c) and (g), which correspond to ${|I|\approx 0}$, where streamlines are parallel to lines of constant vorticity. 
While these lines are approximately straight throughout the R1 roll, they have bent in the anticlockwise direction in the centre of the R2 roll, owing to the increased effects of inertia.
We will subsequently use these qualitative differences to classify rolls by eye rather than computing the vorticity balance for individual states.

\begin{figure}
	\includegraphics[width=0.98\linewidth]{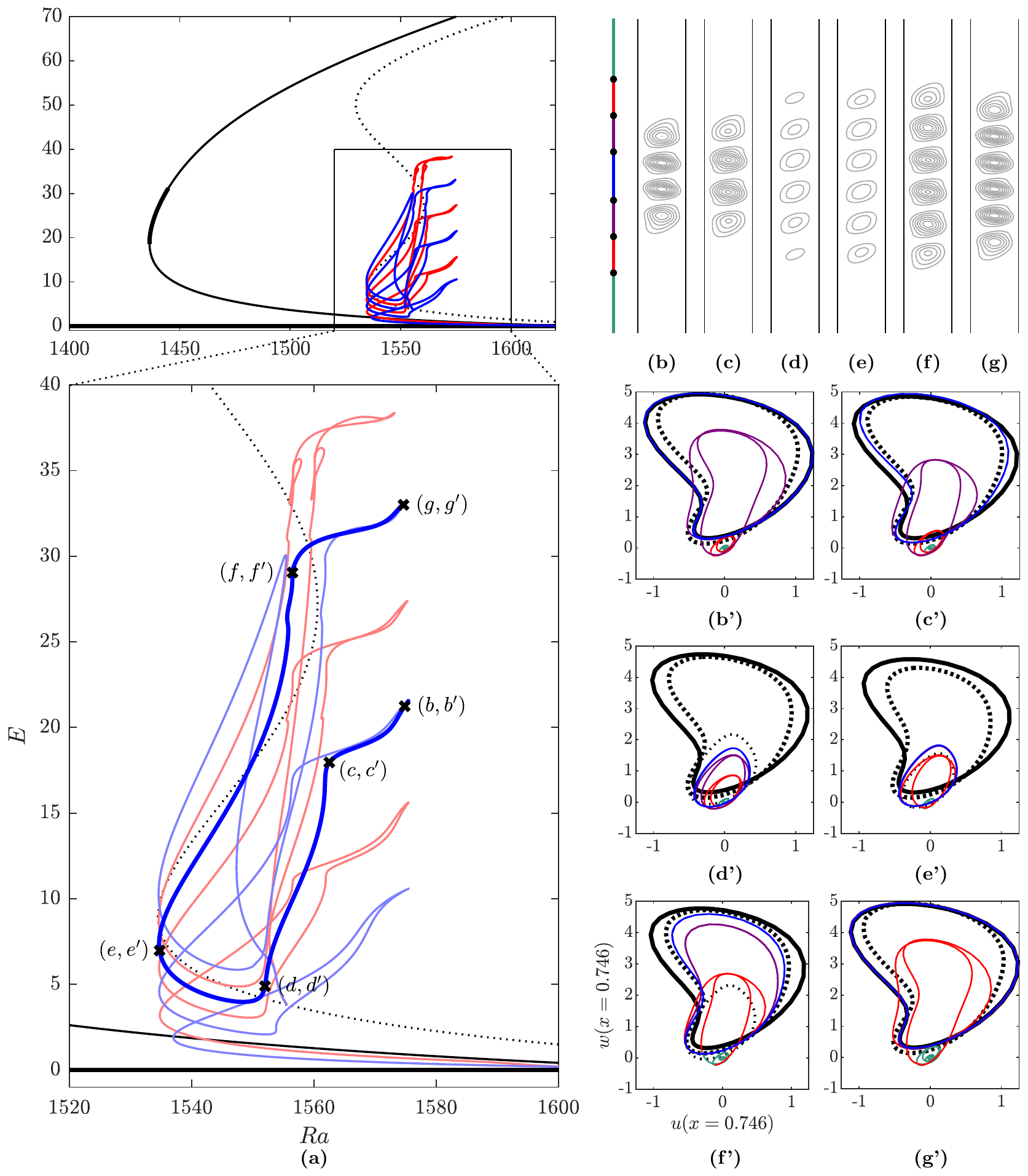}
	\caption[Bifurcation diagram, streamfunction profiles and phase-space representations for ${Pr = 0.12}$]{(a) Bifurcation diagrams for ${Pr = 0.12}$ (Stage~2) showing: the conduction state (thick black solid), P12 (thin black solid), P10 (black dotted), $L^+$ (red solid) and $L^-$ (blue solid).
		Stable (unstable) segments of the primary branches are indicated using bold (thin) lines.
		(b)\textendash{}(g) Streamfunction profiles and (b')\textendash{}(g') corresponding phase-space representations of selected marked states on $L^-$.
		Different types of rolls are represented using different colours in the phase-space plots: strong interior rolls (blue), outer rolls that strengthen to match interior rolls (purple), weak outermost rolls that strengthen over the oscillation (red) and weak-amplitude background rolls (green).
		The bar to the left of panel (b) relates these to the streamfunction profiles.
		Also shown in the phase-space plots are states on the upper branch of P12 (thick black solid) and P10 (thick black dotted) and, where applicable, on the second P10 branch segment (thin black dotted) at the same value of the Rayleigh number.}
	\label{fig:5_Pr012}
\end{figure}

Having observed that the form of the convecton rolls changes along the snaking branches when ${Pr = 0.12}$, we now analyse how and why these changes occur by relating the convectons to the periodic convection states on the primary branches.
To do so, we present streamfunctions and phase-space representations of convectons over a single snaking oscillation for ${Pr = 0.12}$ in figure~\ref{fig:5_Pr012}.
The form of individual rolls varies across each convecton and we have therefore used different colours in the phase-space plots to indicate their form separately.
As pointed out above, the behaviour at the left saddle nodes (e.g., figures~\ref{fig:5_Pr012}(e) and (e')) resembles that observed in Stage~1, where convectons consist of weak R1 rolls. 
These rolls (indicated by the red and indistinguishable blue and purple trajectories in the phase-space plot (e')) most closely resemble those in states on the second branch segment of P10 at the same Rayleigh number (thin black dotted trajectory).
Rolls continue to have this form as the branches are followed towards higher Rayleigh numbers. 
This changes, however, around $Ra\approx 1555$ where we see a near vertical increase in both the kinetic energy of the states (figure~\ref{fig:5_Pr012}(a)) and contribution from inertia in the vorticity equation (figure~\ref{fig:5_vortPr012}(i)).
This corresponds to a rapid strengthening of the central rolls, which also change in form from R1 rolls close to those on the second primary branch segment, to R2 rolls on the upper primary branch segments. 
This is seen in figures~\ref{fig:5_Pr012}(c') and (f'), where the innermost rolls, represented by the blue trajectories, are similar to rolls in states on one of the upper primary branches, represented by either the thick black dotted or solid trajectories.
Continuing the branch segments to the right saddle nodes (b,b') and (g,g'), we find that the size of the central rolls decreases so that the trajectories, shown in blue, are nearly identical to those on the upper branch of P12 (thick black line).
Thus, towards the right edge of the pinning region, when the second branch segment of P10 does not exist, the central rolls in the convectons are R2 rolls.
We further see that the homoclinic orbits of the stable conduction state associated with these localised states being homoclinic orbits between now approach the periodic orbit of a state on an upper primary branch.

\begin{figure}
	\centering
	\includegraphics[width=0.99\linewidth]{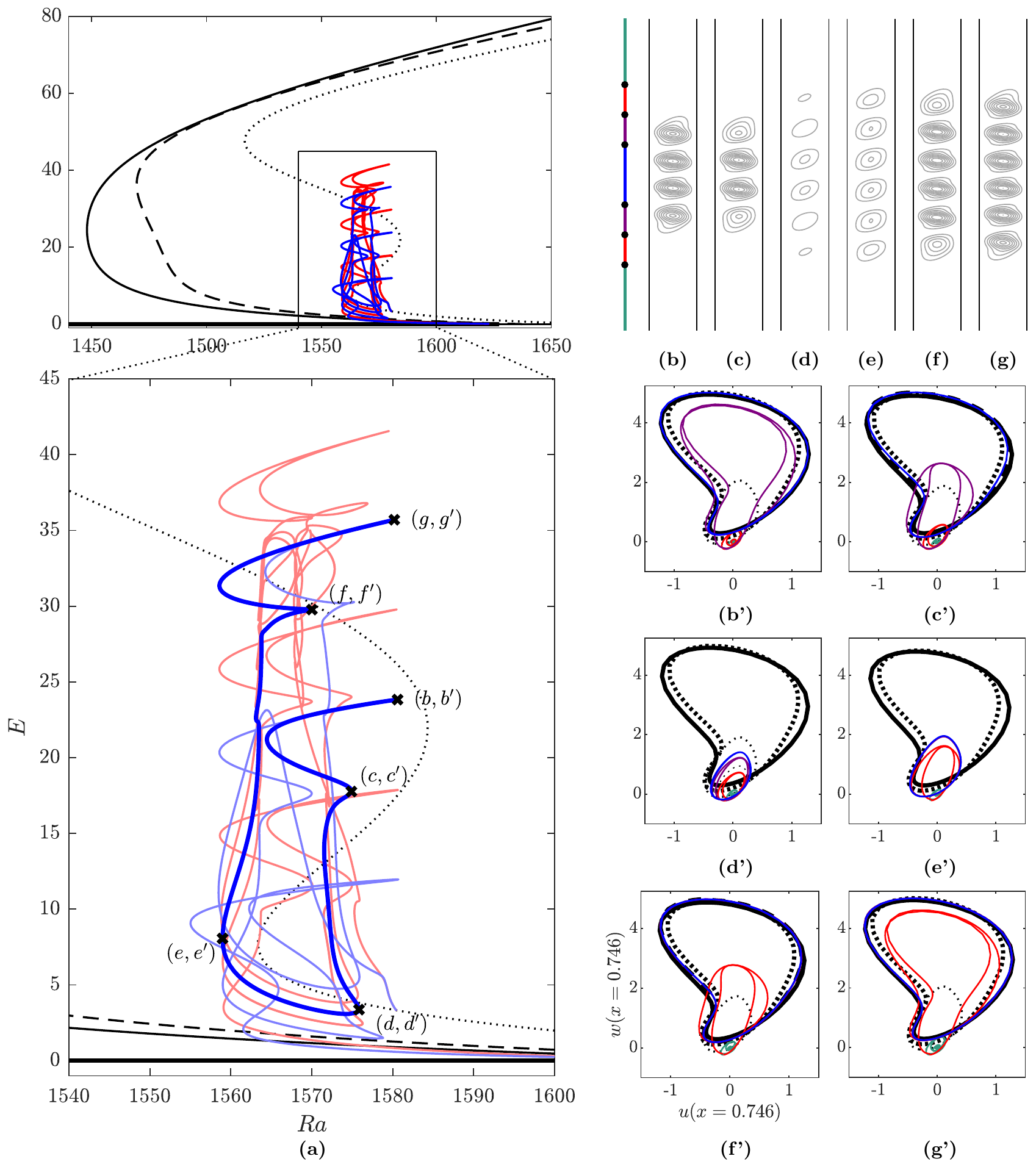}
	\caption[Bifurcation diagram, streamfunction profiles and phase-space representations for ${Pr = 0.11}$]{(a) Bifurcation diagrams for for ${Pr = 0.11}$ (Stage~2), showing: the conduction state (thick black solid), P12 (thin black solid), P11 (black dashed), P10 (black dotted), $L^+$ (red solid) and $L^-$ (blue solid).			
		(b)\textendash{}(g) Streamfunction profiles and (b')\textendash{}(g') corresponding phase-space representations of selected marked states on $L^-$.
		Different types of rolls are represented using different colours in the phase-space plots: strong interior rolls (blue), outer rolls that strengthen to match interior rolls (purple), weak outermost rolls that strengthen over the oscillation (red) and weak-amplitude background rolls (green).
		The bar to the left of panel (b) relates these to the streamfunction profiles.
		Also shown in the phase-space plots are states at the same value of the Rayleigh number on the upper branch of P12 (thick black solid) and P10 (thick black dotted) and, where applicable, on the upper branch of P11 (thick black dashed) and on the first and second P10 branch segments (thin black dotted).
	}
	\label{fig:5_Pr011}
\end{figure}

The steep branch segments near the right saddle nodes (e.g., near (b) and (g) in figure~\ref{fig:5_Pr012}(a)) steepen as the Prandtl number decreases until each segment undergoes a cusp bifurcation just below ${Pr = 0.12}$. 
This leads to the emergence of additional saddle nodes, which provide the hook-like structure of the branches at ${Pr = 0.115}$ (fourth column of figure~\ref{fig:5_bd2lz12le5pr011015}).
The newly formed left saddle node from each pair moves towards lower Rayleigh numbers as the Prandtl number decreases, which increases the multiplicity of convectons within the pinning region.
This multiplicity is exhibited in figure~\ref{fig:5_Pr011} for ${Pr=0.11}$, where we see convectons consisting of weak R1 rolls and those with stronger R2 rolls coexisting over the range of Rayleigh numbers ${1559 \lesssim Ra \lesssim 1576}$.

The transition between the two types of convectons is accompanied with the increasing complexity of the structure of a single snaking oscillation, like the one shown in bold in figure~\ref{fig:5_Pr011}(a).
To interpret the structure of this snaking oscillation, one may consider the oscillation as a combination of three sets of branch segments according to the form of convectons that lie on them.
The first set of branch segments, (b,b')\textendash{}(c,c') and (f,f')\textendash{}(g,g'), originates from the cusp bifurcations near the right edge of the pinning region and, consequently, contains convectons with strong R2 central rolls.
The form of these central rolls most closely resembles rolls in states on the upper branch of P12, which is seen in the phase-space representations at the right saddle nodes (b') and (g') by the blue trajectories (central rolls) following the thick black solid trajectories (upper branch of P12). 
The preferred wavelength of the central rolls increases as each branch segment is followed to either (c) or (f).
This is seen through the blue trajectories in phase space most closely resembling trajectories that correspond to states on the upper branch of P11 (thick black dashed lines in (c') and (f')).
The branch segment (d,d')\textendash{}(e,e') contains convectons with weak R1 central rolls, which are most similar in form to rolls in states on the lower branch segments of P10.
This is evidenced in figures~\ref{fig:5_Pr011}(d') and (g'), by the small, nearly elliptical trajectories associated with convectons that approach the thin black dotted lines, which are associated with states on the second P10 branch segment.
We should note, however, that convectons containing these weak R1 rolls can be found at lower Rayleigh numbers than the first left saddle node of P10 and, thus, in the absence of coexistence with the related spatially periodic states.
The remaining pair of branch segments, (c,c')\textendash{}(d,d') and (e,e')\textendash{}(f,f'), correspond to convectons transitioning between the two regimes with convecton rolls weakening and changing form from R2 to R1, or strengthening and changing form from R1 to R2, respectively.

\subsection{Stage 3: Imperfect bifurcations}
\begin{figure}
	\centering
	\includegraphics[width=0.99\linewidth]{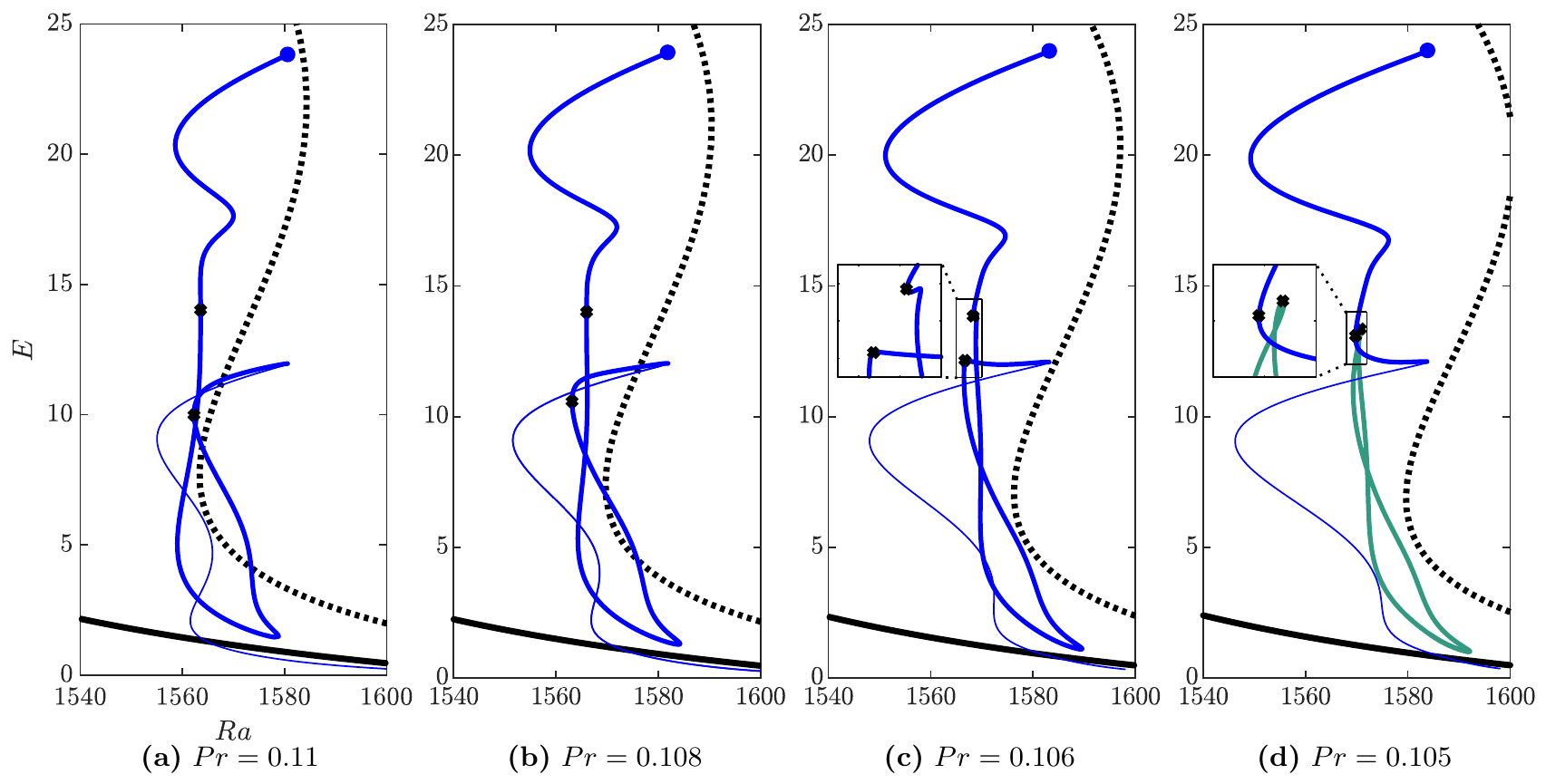}
	\caption[Bifurcation diagrams illustrating part of the breakup of the snaking $L^-$ branch into a main branch and isolas]{Bifurcation diagrams illustrating part of the breakup of the snaking $L^-$ branch (blue) into a main branch (blue) and isolas (green) for (a) ${Pr = 0.11}$, (b) ${Pr = 0.108}$, (c) ${Pr = 0.106}$ and (d) ${Pr = 0.105}$ (Stage~3).
		The primary P12 (black solid) and P10 (black dotted) branches are also shown.
		The connected branch segment between the right saddle nodes corresponding to two- or four-roll convectons and the segments that it breaks into are marked in bold for clarity.}
	\label{fig:5_breakup}
\end{figure}

The first two saddle nodes of P10 continue to move towards larger values of the Rayleigh number as $Pr$ decreases, as shown in figure~\ref{fig:5_breakup}.
By ${Pr = 0.102}$, the first of these saddle nodes has moved past the right edge of the pinning region. 
This leads to each of the snaking secondary branches breaking into a main branch and a number of isolas. 
The two types of convectons---those with R1 rolls (e.g., figures~\ref{fig:5_Pr011}(f) and (g)) and those with R2 rolls (e.g., figures~\ref{fig:5_Pr011}(b) and (j))---now lie on distinct branch segments as a result of this disconnection.
We characterise this transition between ${Pr \approx 0.11}$ and ${Pr \approx 0.102}$ as Stage 3.
The panels in figure~\ref{fig:5_breakup} detail this process by showing how $L^-$ breaks up between the right saddle nodes corresponding to the two- and four-roll convectons.
To aid the reader in following the process, we have marked two points that approach each other, connect and later separate as the Prandtl number decreases.
Between ${Pr = 0.11}$ (figure~\ref{fig:5_breakup}(a)) and ${Pr \approx 0.106}$ (figure~\ref{fig:5_breakup}(c)), the state at the marked left saddle node increases in amplitude, while the second state approaches this saddle node as the near-vertical return branch segment pinches away towards lower Rayleigh numbers (see inset of figure~\ref{fig:5_breakup}(c)).
These states proceed to connect at some point between ${Pr = 0.106}$ and ${Pr = 0.105}$ in a transcritical bifurcation, before later separating as $Pr$ decreases.
This imperfect bifurcation results in an isola (shown in green in figure~\ref{fig:5_breakup}(d)) disconnecting from the main branch.

The forms of both the convectons that lie on this isola and those that remain on the main branch may be deduced by first relating the isola and main branch when ${Pr = 0.105}$ (figure~\ref{fig:5_breakup}(d)) back to the corresponding segments of the following snaking oscillation when ${Pr = 0.11}$ (figure~\ref{fig:5_breakup}(a)).
In doing so, we find that convectons on the isola originated from states on the branch segment between the marked points, which corresponds to convectons with R1 rolls and the transition between R1 and R2 rolls, as was seen between points (c,c') and (f,f') in figure~\ref{fig:5_Pr011}.
In contrast, convectons remaining on the main branch originated from other sections of the ${Pr=0.11}$ snaking oscillation, where the central rolls resembled those on the upper primary branch segments.
This breakup process therefore separates convectons with R1 rolls, which now lie on the isola, from convectons with smaller and stronger R2 rolls, which remain on the main branch.

The following snaking oscillations of $L^-$ and those on $L^+$ undergo similar, albeit more complicated, breakup processes.
For each oscillation, analogous points to those marked in figure~\ref{fig:5_breakup} connect in multiple stages, which result in several isolas disconnecting from the main branch.
We will not discuss the details of these subsequent connections here.

\begin{figure}
	\centering
	\includegraphics[width=0.99\linewidth]{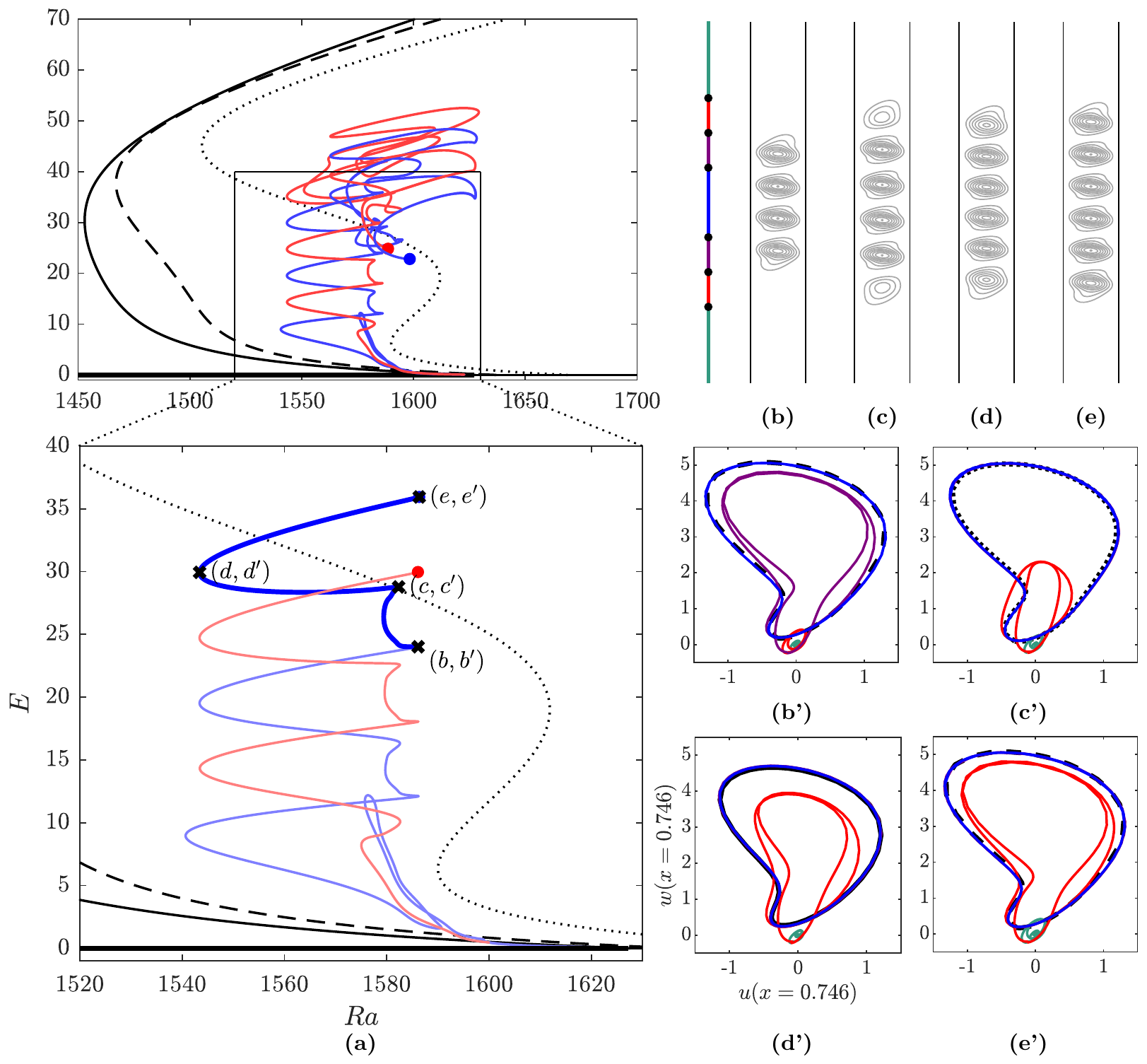}
	\caption[Bifurcation diagram, streamfunction profiles and phase-space representations for ${Pr = 0.102}$]{(a) Bifurcation diagrams when ${Pr = 0.102}$ (Stage~3) showing: the conduction state (thick black solid), P12 (thin black solid), P11 (black dashed), P10 (black dotted), $L^+$ (red solid) and $L^-$ (blue solid).
		In the upper panel, $L^-$ and $L^+$ are terminated at the blue and red points, whereas they are terminated at the marked right saddle nodes below ${E = 40}$ in the lower panel for clarity.
		(b)\textendash{}(g) Streamfunction profiles and (b')\textendash{}(g') corresponding phase-space representations of selected marked states on $L^-$.
		In (b')\textendash{}(g'), different types of rolls are represented using different colours in the phase-space plots: strong interior rolls (blue), outer rolls that strengthen to match interior rolls (purple), weak outermost rolls that strengthen over the oscillation (red) and weak-amplitude background rolls (green).
		The bar to the left of panel (b) relates these to the streamfunction profiles.
		Also shown in the phase-space plots are states on the upper segment of the primary branch whose rolls at the same value of the Rayleigh number are most similar to the interior convecton rolls, again using the convention: P12 (thick black solid), P11(thick black dashed) and P10 (thick black dotted).}
	\label{fig:5_Pr0102solbranch}
\end{figure}

This breakup process ultimately leads to convectons lying on one of the two main branches that display different snaking behaviour to that seen at higher Prandtl numbers (e.g., see figure~\ref{fig:5_Pr0102solbranch}(a) for ${Pr=0.102}$) or on one of a number of isolas that are not shown in figure~\ref{fig:5_Pr0102solbranch}(a), to avoid cluttering.
We first note that the isola seen in figure~\ref{fig:5_breakup}(d) when ${Pr=0.105}$ reconnects to $L^-$ at small amplitude and is responsible for the initial excursion to ${E\approx 12}$ that the main snaking branch exhibits for $Pr = 0.102$ before returning to small amplitudes.
The branch then proceeds to snake upwards, with each snaking oscillation being composed of two parts: a small section (e.g., between the states labelled (b,b') and those labelled (c,c') on figure \ref{fig:5_Pr0102solbranch}), and a larger section (e.g., between the states labelled (c,c') and those labelled (e,e')).
The small section is associated with the nucleation and strengthening of a pair of outer R1 rolls, which is exemplified by the red trajectories in (b') and (c').
Meanwhile, the outer central rolls (purple trajectories) strengthen to match the inner central rolls (blue trajectories), so that the blue and purple trajectories are indistinguishable by the right saddle node (c').
These central R2 rolls also undergo changes over this small segment: they initially resemble rolls in states on the upper branch of P11 (thick black dashed trajectory in (b')) to finally resemble those on the upper branch of P10 (thick black dotted trajectory in (c')).
The larger section part of the snaking oscillation (between (c,c') and (e,e')) is predominantly associated with the strengthening and, eventually, transition to R2 of the outermost R1 rolls that were nucleated during the first section of the snaking oscillation.
The four central rolls additionally change from resembling P10 rolls at (c,c') to P12 rolls at the following left saddle node (d,d') before returning to P11 rolls at the right saddle node (e,e').
Consequently, the net effect of a complete snaking oscillation is for the convecton to extend in length by two central R2 rolls. 

\begin{figure}
	\centering
	\includegraphics[width=0.99\linewidth]{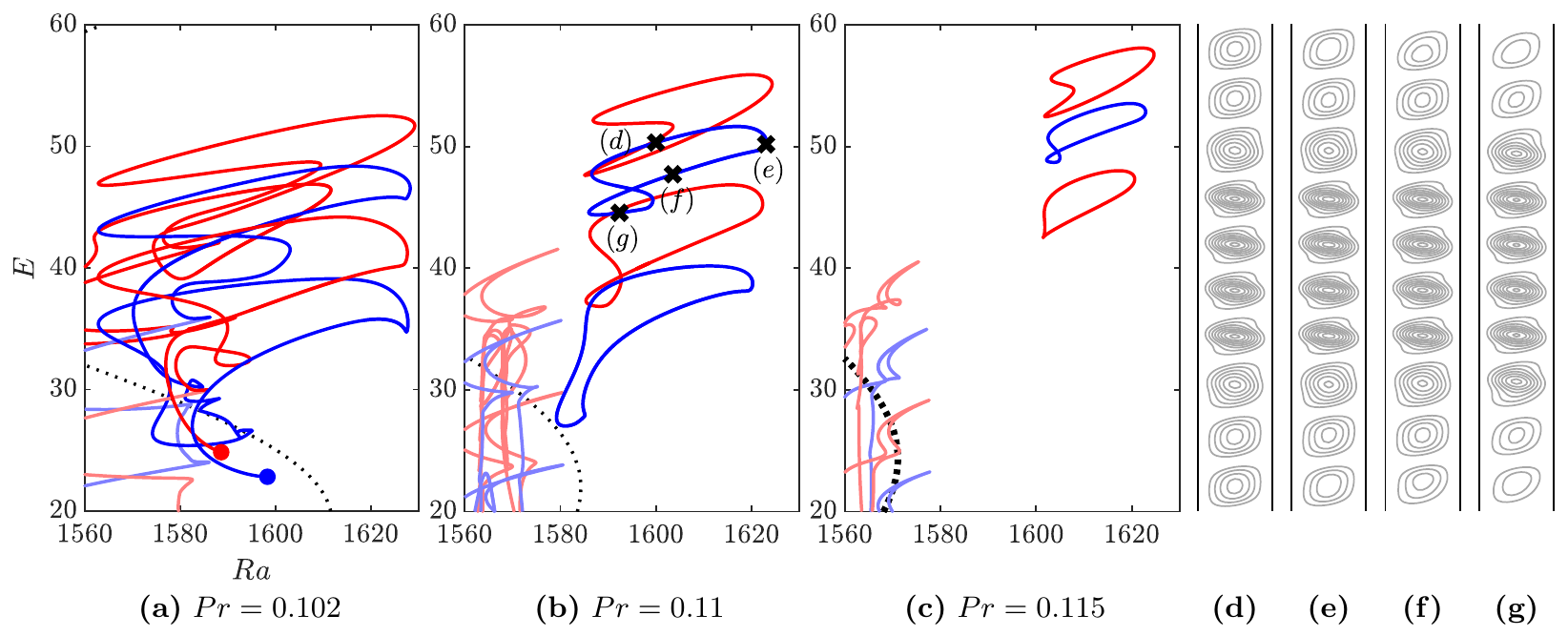}
	\caption[Branch segments and profiles of non-uniform, domain-filling patterned states]{Branch segments and profiles of non-uniform, domain-filling patterned states (Stage~3).
		(a) Extension of the bifurcation diagram shown in figure~\ref{fig:5_Pr0102solbranch}(a) for ${Pr = 0.102}$, again with branches terminated at the marked points for clarity.
		(b) Snaking $L^+$ and $L^-$ branches for ${Pr = 0.11}$ with four isolas, which correspond to states with two (lower blue), three (lower red), four (upper blue) or five (upper red) stronger interior rolls in a background of weaker rolls.
		(c) Similarly to (b), except for ${Pr = 0.115}$, where only the isolas with three or four stronger central rolls are found.
		(d)\textendash{}(g) Streamfunction profiles for the marked states on the four strong roll isola when ${Pr = 0.11}$.}
	\label{fig:5_bdlz12le5pr0102isolajoined}
\end{figure}
This kind of snaking stops when $L^+$ reaches a five-roll state and $L^-$ reaches a six-roll state, marked by the dots in figure~\ref{fig:5_Pr0102solbranch}(a).
The branches instead exhibit different behaviour, magnified in figure~\ref{fig:5_bdlz12le5pr0102isolajoined}(a), that is bounded between ${Ra \approx 1563}$ and ${Ra \approx 1629}$, which we note is close to the critical Rayleigh number ${Ra_c\approx 1627}$.
These oscillations correspond to domain-filling patterned states with ten rolls, similar to those in figures~\ref{fig:5_bdlz12le5pr0102isolajoined}(d)\textendash{}(g), where central R2 rolls are embedded in a background of weaker R1 rolls.
Tracking these states into higher Prandtl numbers, we find that they lie on isolas that are disconnected from both each other and the main snaking branches (see figure~\ref{fig:5_bdlz12le5pr0102isolajoined}(b)).
Along an isola, the number of stronger central rolls remains the same, while the background rolls change from having a near-uniform amplitude (e.g., figure~\ref{fig:5_bdlz12le5pr0102isolajoined}(d)) to displaying a larger variation in amplitude (e.g., figure~\ref{fig:5_bdlz12le5pr0102isolajoined}(g)).
The isolas become smaller as the Prandtl number increases, but at different rates, so that by ${Pr = 0.115}$ (figure~\ref{fig:5_bdlz12le5pr0102isolajoined}(c)), the isola containing the patterned states with two strong central rolls no longer exists, while the one containing states with four central rolls persists until about ${Pr \approx 0.1191}$.

\begin{figure}
	\centering
	\includegraphics[width=0.99\linewidth]{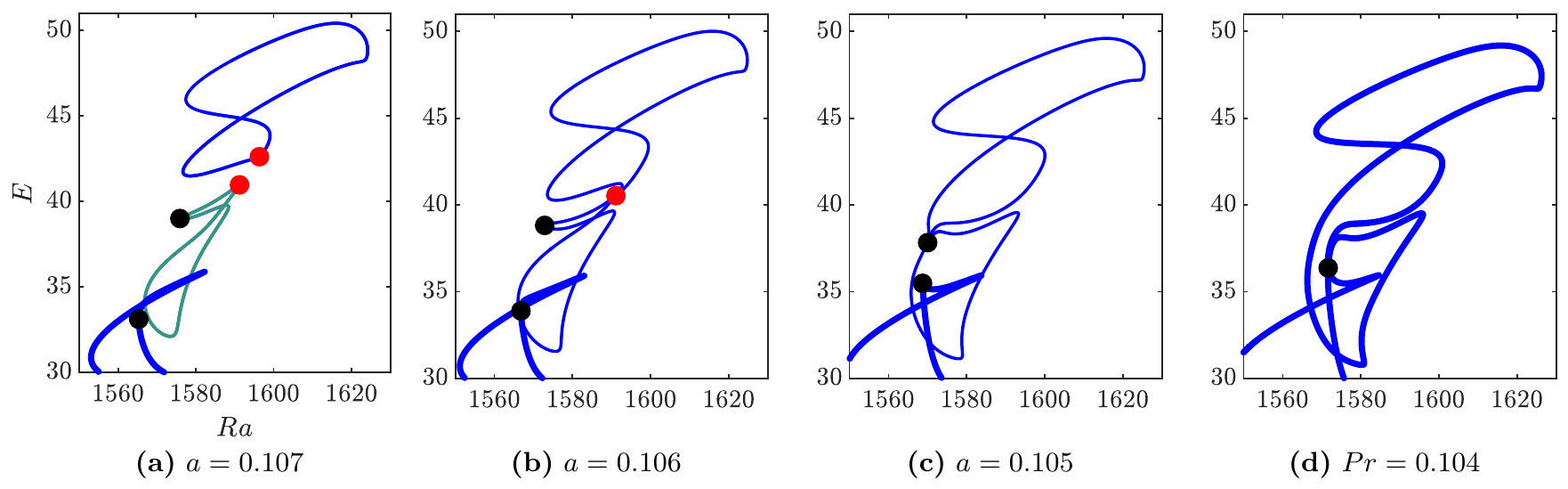}
	\caption[Bifurcation diagrams showing how the isolas of domain-filling patterned states with stronger central rolls connect to $L^-$ ]{Bifurcation diagrams showing how the isolas of domain-filling patterned states with stronger central rolls connect to $L^-$ for (a) ${Pr = 0.107}$, (b) ${Pr = 0.106}$, (c) ${Pr = 0.105}$ and (d) ${Pr = 0.104}$ (Stage~3).
		The main branch $L^-$ is shown by the thick blue line. 
		The isolas originating from that marked in figure~\ref{fig:5_bdlz12le5pr0102isolajoined}(b) are shown by the thin blue lines.
		A second isola that connects to this isola at ${Pr \approx 0.106}$ is shown in green.
		The red and black dots mark points that successively connect as the Prandtl number decreases.}
	\label{fig:5_bdisolajoin}
\end{figure}
Figure~\ref{fig:5_bdisolajoin} depicts how the isola with four stronger central rolls that was marked in figure~\ref{fig:5_bdlz12le5pr0102isolajoined}(b) connects to $L^-$ as the Prandtl number decreases from ${Pr=0.107}$ (figure~\ref{fig:5_bdisolajoin}(a)) to ${Pr=0.104}$ (figure~\ref{fig:5_bdisolajoin}(d)).
We first find that this isola connects to the top right section of a second isola (green in figure~\ref{fig:5_bdisolajoin}(a)) of lower energy states, whose outer rolls are amplitude-modulated in a similar way to those shown in figure~\ref{fig:5_bdlz12le5pr0102isolajoined}(g).
This connection is illustrated in figures~\ref{fig:5_bdisolajoin}(a) and (b) by the two red dots at ${Pr = 0.107}$ meeting by ${Pr = 0.106}$.
The isola resulting from this merger proceeds to connect to $L^-$ between ${Pr = 0.105}$ (figure~\ref{fig:5_bdisolajoin}(c)) and ${Pr = 0.104}$ (figure~\ref{fig:5_bdisolajoin}(d)).
This is shown by the black dots in figure~\ref{fig:5_bdisolajoin}, as the point that was originally a left saddle node of the lower isola at ${Pr = 0.107}$ moves to lower energy and towards the pinch-off point of $L^-$ seen at ${Pr = 0.105}$ (figure~\ref{fig:5_bdisolajoin}(c)). 
The two points connect by ${Pr = 0.104}$, so, after reaching the six-roll convectons, the $L^-$ branch heads towards larger energy and follows the path of the isola, before heading towards lower energy as it does at higher Prandtl numbers.
At the lower-$Pr$ limit of Stage 3, the branches $L^+$ and $L^-$ therefore contain both convectons and domain-filling patterned states.

The subsequent behaviour of the branches $L^+$ and $L^-$ after these oscillations, i.e., beyond the blue and red dots in the top panel of figure~\ref{fig:5_bdisolajoin}(a), is unclear.
They display no organised structure, nor did we manage to continue them to a termination point.
As the states they carry are close to being domain-filling, the structure of these branches are sensitive to the domain size and they are not as relevant to the study of localised states as the convectons of smaller extent described previously.
We have therefore not tried to characterise them any further.

\clearpage
\subsection{Stage 4: Disconnected convecton branches and small-amplitude snaking}

In the previous stage, we saw how the pair of rolls that nucleate at the right saddle nodes (e.g., figure~\ref{fig:5_Pr0102solbranch}(b)) take a similar form to R1 rolls in states on the lower branch segments of P10.
However, as the Prandtl number continues to decrease, the primary bifurcation of the conduction state becomes increasingly supercritical and the first two saddle nodes of the P10 branch continue to move towards higher Rayleigh numbers, further away from the right edge of the pinning region.
This impacts the snaking in two main ways.
Firstly, each snaking branch breaks up into two snaking branches: one with small-amplitude convectons and one with large-amplitude convectons.
Secondly, the branch segments between convectons with $n$ and $n+2$ rolls disconnect, with convectons on these branches instead continuously transitioning into domain-filling states similar to those discussed at the end of the previous section.
We will describe these two changes in this section and refer to them both as Stage~4.

\subsubsection{Large-amplitude convectons}
\begin{figure}
	\centering
	\includegraphics[width=0.95\linewidth]{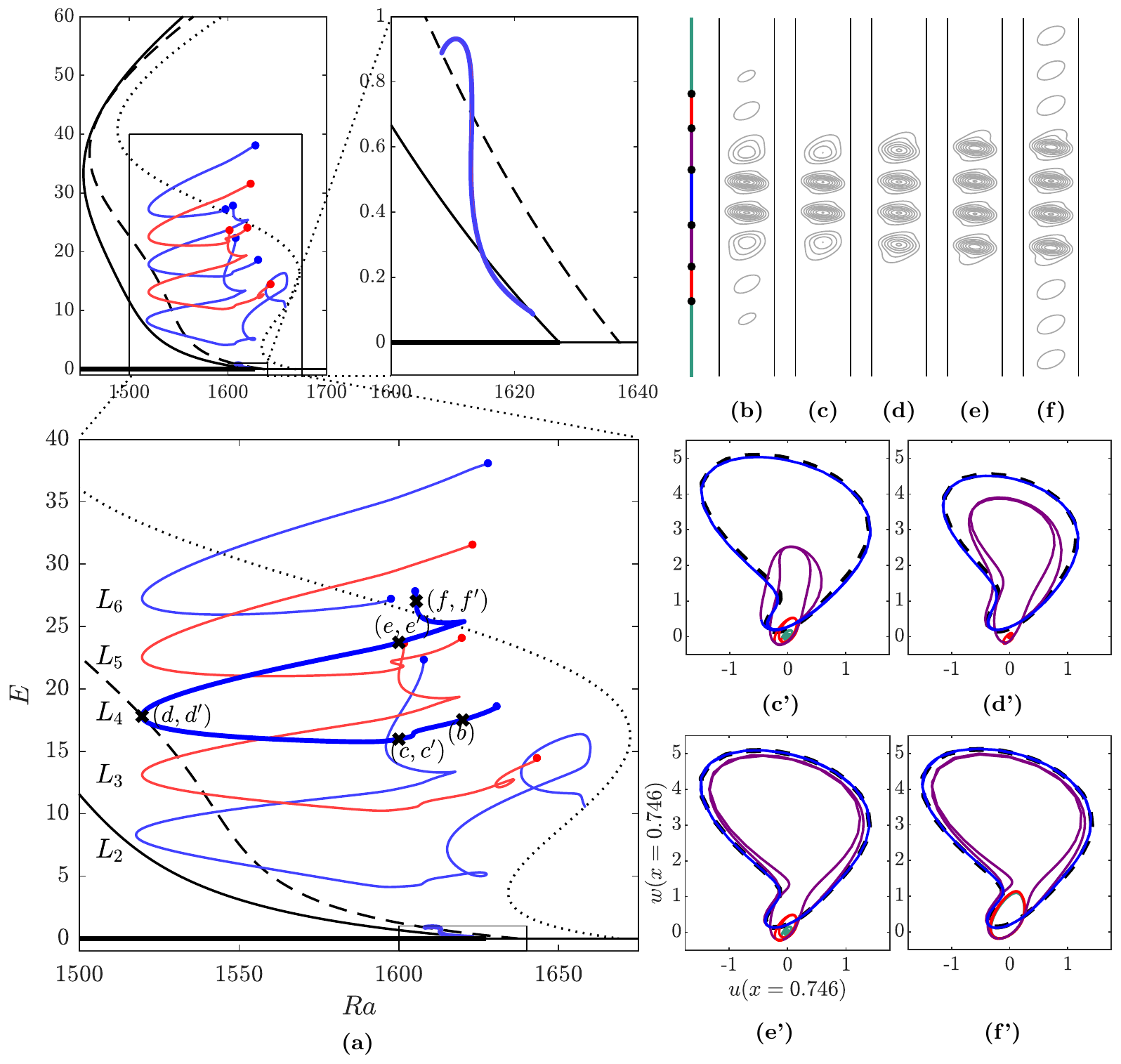}
	\caption[Bifurcation diagram, streamfunction profiles and phase-space representations for ${Pr = 0.09}$]{(a) Bifurcation diagrams for ${Pr = 0.09}$ (Stage~4). Top left: zoomed out bifurcation diagram. Top right: magnification around the secondary bifurcation of P12. 
		The branches shown in (a) are: conduction state and P12 (black solid), P11 (black dashed), P10 (black dotted) and $L_i$, branch of convectons with $i$ rolls (blue (red) when $i$ is even (odd)).
		The branches $L_i$ have been terminated at the marked blue and red dots where the branches subsequently head back towards lower Rayleigh numbers, for clarity. 
		(b)\textendash{}(f) Streamfunctions of the labelled states of $L_4$ and (c'), (d'), (e') and (f') corresponding phase-space representations for four of these states.
		The phase-space plots show the convectons on $L_4$ (blue) and the state on the upper branch of P11 (black dashed) at the same value of the Rayleigh number.}
	\label{fig:5_bdlz12le5pr009}
\end{figure}

Figure~\ref{fig:5_bdlz12le5pr009} presents the bifurcation diagram for ${Pr=0.09}$, which demonstrates the behaviour after the changes within this stage have occurred.
In particular, we note the presence of a pair of small-amplitude snaking branches that bifurcates from P12 and terminates on P11 and of five disconnected branch segments of large-amplitude convectons.
We introduce the notation $L_i$ to label these new convecton branches, where $i$ identifies the number of rolls contained in a given branch convecton for subcritical Rayleigh numbers ${Ra<Ra_c}$.
While these branches are now disconnected from each other, the organised nature of the convecton branches for subcritical Rayleigh numbers persists and the left saddle nodes vertically align around ${Ra\approx 1520}$.

Changes along the organised branch segments can be understood by following each branch segment $L_i$ from ${Ra\approx 1600}$ on the lower branch segment toward the upper branch segment.
Considering the branch of four-roll convectons $L_4$, for example, we find that the convecton at ${Ra\approx 1600}$ on the lower branch segment (figure~\ref{fig:5_bdlz12le5pr009}(c,c')) contains two inner R2 rolls (blue trajectories in (c')) whose form closely follows that in states on the upper branch of P11 (thick black dashed trajectory in (c')) and two outer R1 rolls (purple trajectories in (c')).
The inner convecton rolls continue to follow those on the upper branch of P11 at the same Rayleigh number as $L_4$ is followed firstly towards the left saddle node at ${Ra\approx 1520}$ (d,d') and later towards higher Rayleigh numbers (e,e').
We should note that this differs from the results at ${Pr=0.102}$, where the inner rolls changed from resembling rolls on the upper branch segments of P10 (figure~\ref{fig:5_Pr0102solbranch}(c')) to P12 (figure~\ref{fig:5_Pr0102solbranch}(d')) and then to P11 (figure~\ref{fig:5_Pr0102solbranch}(e')) over the corresponding oscillation.
Meanwhile, the two outer rolls undergo a similar increase in amplitude and change in structure to that seen when ${Pr=0.102}$, in that they change from R1 rolls on the lower branch segment (figure~\ref{fig:5_bdlz12le5pr009}(c')) to R2 rolls at the left saddle node (figure~\ref{fig:5_bdlz12le5pr009}(d')), before increasing in amplitude as the branch is followed towards higher Rayleigh numbers (figure~\ref{fig:5_bdlz12le5pr009}(e')).

Continuing each upper branch segment of $L_i$ from ${Ra\approx 1600}$ towards larger kinetic energy (e.g., figure~\ref{fig:5_bdlz12le5pr009}(e,e') to (f) and beyond), we find that the background conduction state fills with weak rolls that are uniform in amplitude, instead of nucleating a pair of rolls outside of the existing convecton as was seen at higher Prandtl numbers.
The subsequent steady states are thus domain-filling states with $i$ strong interior rolls within a background of weaker rolls, as illustrated in figure~\ref{fig:5_bdlz12le5pr009}(f).
Following this background nucleation, the branches exhibit complex, seemingly unstructured trajectories between ${Ra\approx 1520}$ and ${Ra\approx 1650}$, which we did not aim to characterise.
These trajectories are not indicated in figure~\ref{fig:5_bdlz12le5pr009}(a), where each branch segment is truncated at right saddle-nodes, marked by the blue and red dots, to keep the figure readable.

Contrasting behaviour arises as each lower branch segment of $L_i$ is continued towards higher Rayleigh numbers (e.g., figure~\ref{fig:5_bdlz12le5pr009}(c,c') to (b) and beyond), as we find that weak R1 rolls successively nucleate and strengthen outside of the stronger central R2 rolls.
This strengthening is non-uniform and rather occurs in a spatially modulated manner so that the amplitude of these background rolls decreases outwards from the central rolls, as seen in figure~\ref{fig:5_bdlz12le5pr009}(b).
Along $L_2$, the branch of two-roll convectons (lowest blue branch in figure~\ref{fig:5_bdlz12le5pr009}(a)), this modulation is accompanied by the inner rolls changing to resemble rolls on the second branch segment of P10 and the branch is seen to bifurcate from a modulational instability of P10 around ${Ra\approx1658}$.
We were unable to determine the origin of the other branches of convectons seen in figure~\ref{fig:5_bdlz12le5pr009}(a).
However, we found that they do not bifurcate from the first secondary bifurcation of P12, as was the case when ${Pr = 0.102}$, since the pair of secondary branches that instead bifurcate from that point undergo small-amplitude snaking over a narrow range of Rayleigh numbers and terminate on P11.

\subsubsection{Small-amplitude convectons}

\begin{figure}
	\centering
	\includegraphics[width=0.99\linewidth]{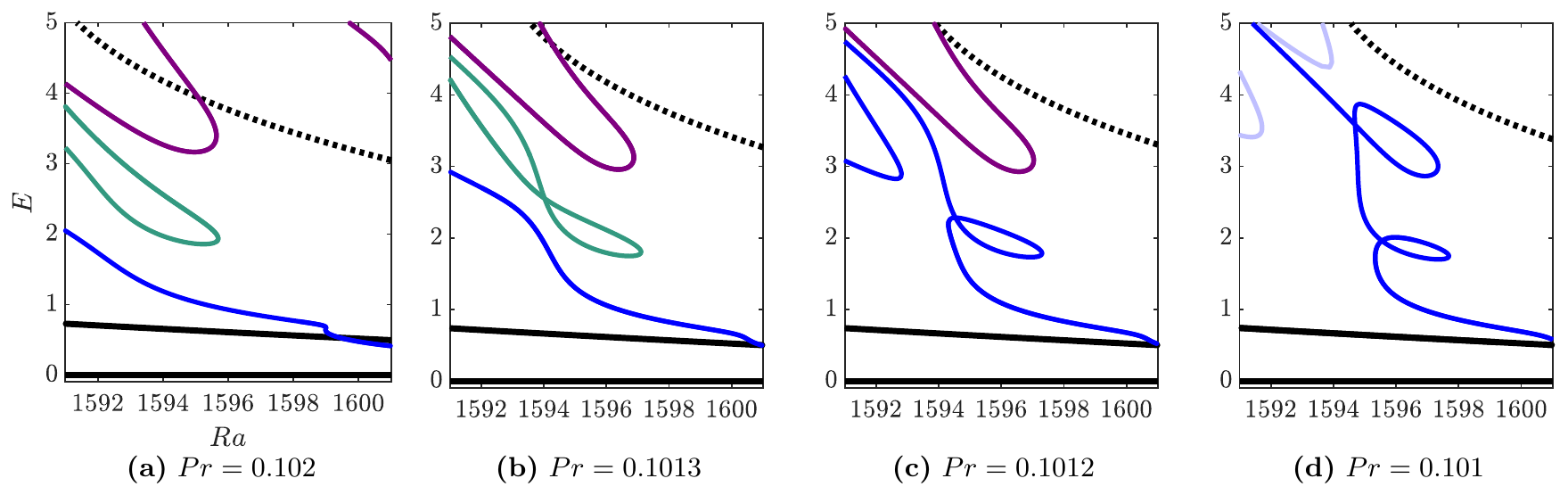}
	\caption[Magnification of bifurcation diagrams showing how the isolas reconnect to the secondary branch during Stage 4]{Magnification of bifurcation diagrams for (a) ${Pr = 0.102}$, (b) ${Pr = 0.1013}$, (c) ${Pr = 0.1012}$ and (d) ${Pr = 0.101}$ (Stage~4) showing how the isolas reconnect to the secondary branch.
		The branches shown are: conduction state (black solid), P12 (black solid), P10 (black dotted), secondary branch $L^-$ that bifurcates from P12 (blue), isolas that break from the snaking $L^-$ branch during Stage 3 (green), a secondary branch containing states with an even number of rolls that bifurcates from a secondary bifurcation of P10 and extends towards large-amplitude (purple) and the branch segment that disconnects from $L^-$ and snakes at large-amplitude (light blue). }
	\label{fig:5_isolasnaking}
\end{figure}
We now turn our attention to both the origin and properties of the small-amplitude snaking that was seen in the top right panel of figure~\ref{fig:5_bdlz12le5pr009}(a).
Parts of the branch segments involved in this snaking come from the lower part of the isolas that disconnected from the main snaking branches in Stage 3. 
This is shown in figure~\ref{fig:5_isolasnaking}, where the isola containing states with six weak R1 rolls (marked in green) undergoes multiple twists (figure~\ref{fig:5_isolasnaking}(b)) before joining to the lower part of the secondary branch (marked in blue) near the first of these crossing points.
This leads to the small loop seen at ${Pr = 0.1012}$ (figure~\ref{fig:5_isolasnaking}(c)) and a subsequent excursion (not shown), where the branch first follows the remaining section of the isola before heading to lower Rayleigh numbers by continuing its original path and snaking at large-amplitude.

Between ${Pr = 0.1012}$ and ${Pr = 0.101}$ (figure~\ref{fig:5_isolasnaking}(d)), the lower part of the secondary branch that bifurcates from a modulational instability of P10 (marked in purple in figure~\ref{fig:5_isolasnaking}(c)) connects with $L^-$. 
This likely occurs by the former branch first twisting over itself to form a second small loop between ${1595<Ra<1597}$ with ${E\approx 3}$.
The lower part of this twisted branch proceeds to merge with the secondary branch bifurcating from P12 (shown in blue) to form the second small loop seen when ${Pr = 0.101}$ (figure~\ref{fig:5_isolasnaking}(d)).
This separates the secondary branch $L^-$ when ${Pr = 0.102}$ (shown in blue) into two: a small-amplitude snaking branch that bifurcates from P12 and terminates on P10 (blue in figure~\ref{fig:5_isolasnaking}(d)), and a branch segment that continues to large amplitude and contains fully localised states similar to those seen in figure~\ref{fig:5_Pr0102solbranch} (shown in light blue in figure~\ref{fig:5_isolasnaking}(d)).		
\begin{figure}
	\centering
	\includegraphics[width=0.99\linewidth]{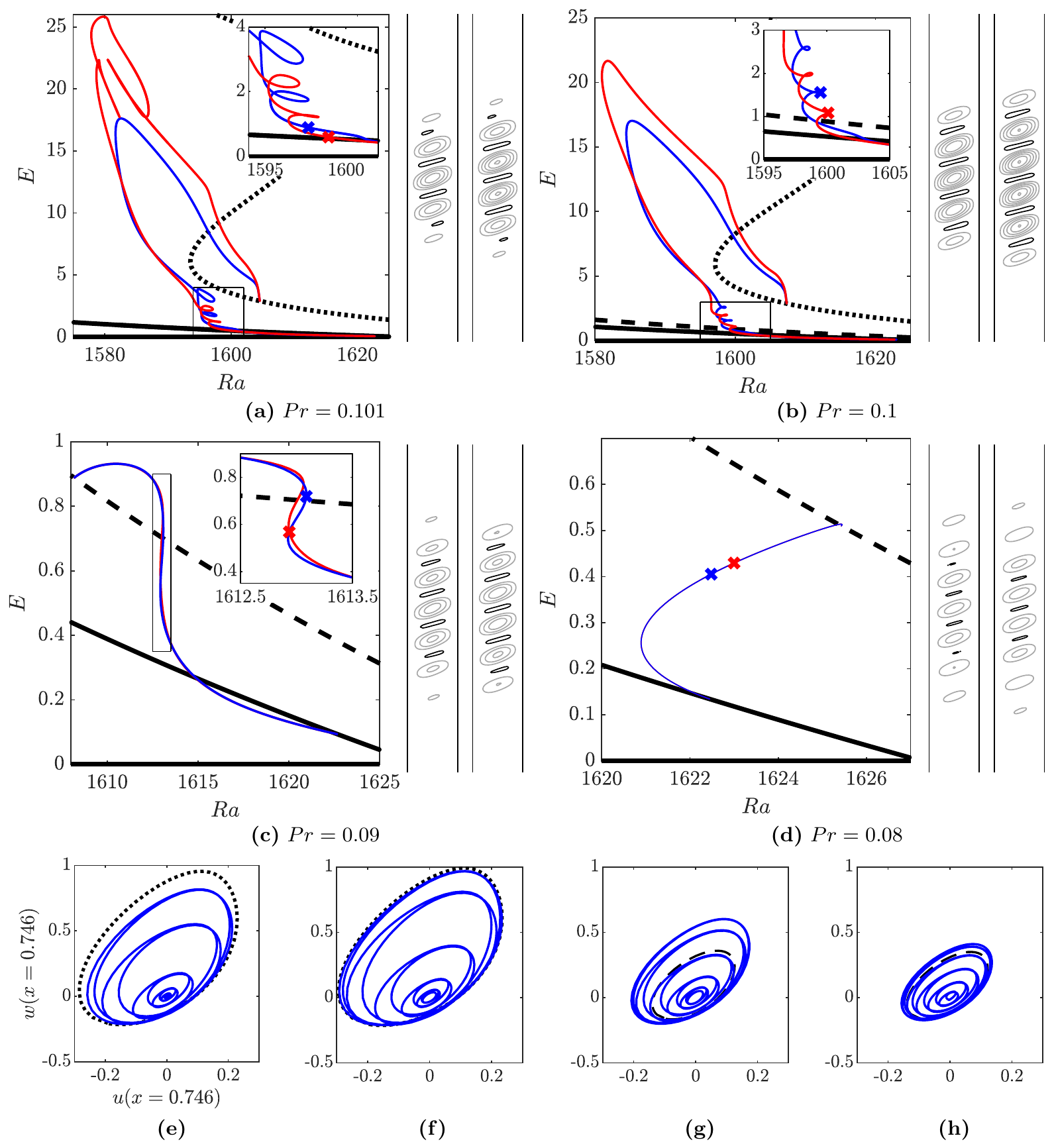}
	\caption[Small-amplitude snaking secondary branches as the Prandtl number decreases within Stage 4]{Small-amplitude snaking secondary branches for (a) ${Pr = 0.101}$, (b) ${Pr=0.1}$, (c) ${Pr=0.09}$ and (d) ${Pr = 0.08}$ (Stage~4).
		The branches shown in the bifurcation diagrams are: conduction state (thick black solid), P12 (thin black solid), P11 (black dashed), P10 (black dotted), $L^+$ (red solid) and $L^-$ (blue solid).	
		The pair of streamfunction profiles indicate the steady states marked on both of the secondary branches (left: $L^+$, right: $L^-$).
		Contour intervals of $0.05$ were used and grey (black) contours represent anticlockwise (clockwise) flow.
		(e)\textendash{}(h) Phase-space plots of the marked small-amplitude states on $L^-$ for (e) ${Pr = 0.101}$, (f) ${Pr = 0.1}$, (g) ${Pr = 0.09}$ and (h) ${Pr = 0.08}$.
		These plots show the trajectories for the convecton on $L^-$ (blue), together with either states on the lower P10 (black dotted in (e,f)) or P11 (black dashed in (g,h)) branch at the same value of the Rayleigh number. }
	\label{fig:5_P12secbif}
\end{figure}
In larger domains, we anticipate that the additional isolas formed during Stage 3 undergo similar twists and connections to the secondary branch, which would increase the number of small loops before the final separation occurs.

States evolve along the resulting narrow snaking branches (shown in figures~\ref{fig:5_P12secbif}(a) and (b) for ${Pr = 0.101}$ and ${Pr = 0.1}$, respectively) in the typical way.
After an initial spatial modulation, states enter the small pinning region with either three or four weak R1 rolls, depending upon the secondary branch.
Additional rolls nucleate on either side of the convecton and proceed to grow in amplitude as the two secondary branches are continued in the direction of increasing energy.
The form of rolls in these small-amplitude convectons is controlled by the lower P10 branch, since the orbits for the two inner rolls (outer blue curves) in an established convecton for ${Pr = 0.1}$ closely follow the orbit for weakest P10 rolls (black dotted curve) in figure~\ref{fig:5_P12secbif}(f).
When ${Pr \geqslant 0.1}$, this snaking is followed by an excursion to larger roll amplitudes, before both secondary branches terminate at a modulational instability of P10 prior to the first left saddle node.

Since the P10 branch becomes increasingly supercritical as the Prandtl number decreases, its first saddle node moves towards larger Rayleigh numbers.
We find that the thin snaking branches cease to terminate on P10 by ${Pr = 0.09}$ and they instead terminate on P11 (see figures~\ref{fig:5_P12secbif}(c) and (d)).
Continuing to decrease the Prandtl number towards $Pr_c$, where the primary bifurcation transitions from being subcritical to supercritical, we observe similar trends to those seen in other approaches to supercriticality in finite domains (e.g., \cite{burke2006localized,kao2012weakly}).
These include the width of the snaking decreasing until the branches become nearly indistinguishable by ${Pr = 0.08}$ (figure~\ref{fig:5_P12secbif}(d)), and saddle nodes colliding in cusp bifurcations from the bottom of the snaking to increase the number of rolls in the convectons.
In an infinite domain, we might thus anticipate that the pinning region for this small-amplitude snaking becomes exponentially thin as ${Pr_c\approx 0.062}$ is approached and that these localised states persist up to this limit.
However, in the 12-wavelength domain considered here, finite-size effects result in the first secondary bifurcation of P12 changing from being stationary at ${Pr = 0.073}$ to oscillatory at ${Pr = 0.072}$, as was seen by the red points in figure~\ref{fig:5_secbifloclz}.
The small-amplitude modulated states seen in figure~\ref{fig:5_P12secbif} thus become time-dependent and we were unable to numerically continue the corresponding branches.

\clearpage
		\subsection{Stage 5: Supercritical convectons}		
\begin{figure}
	\centering
	\includegraphics[width=0.99\linewidth]{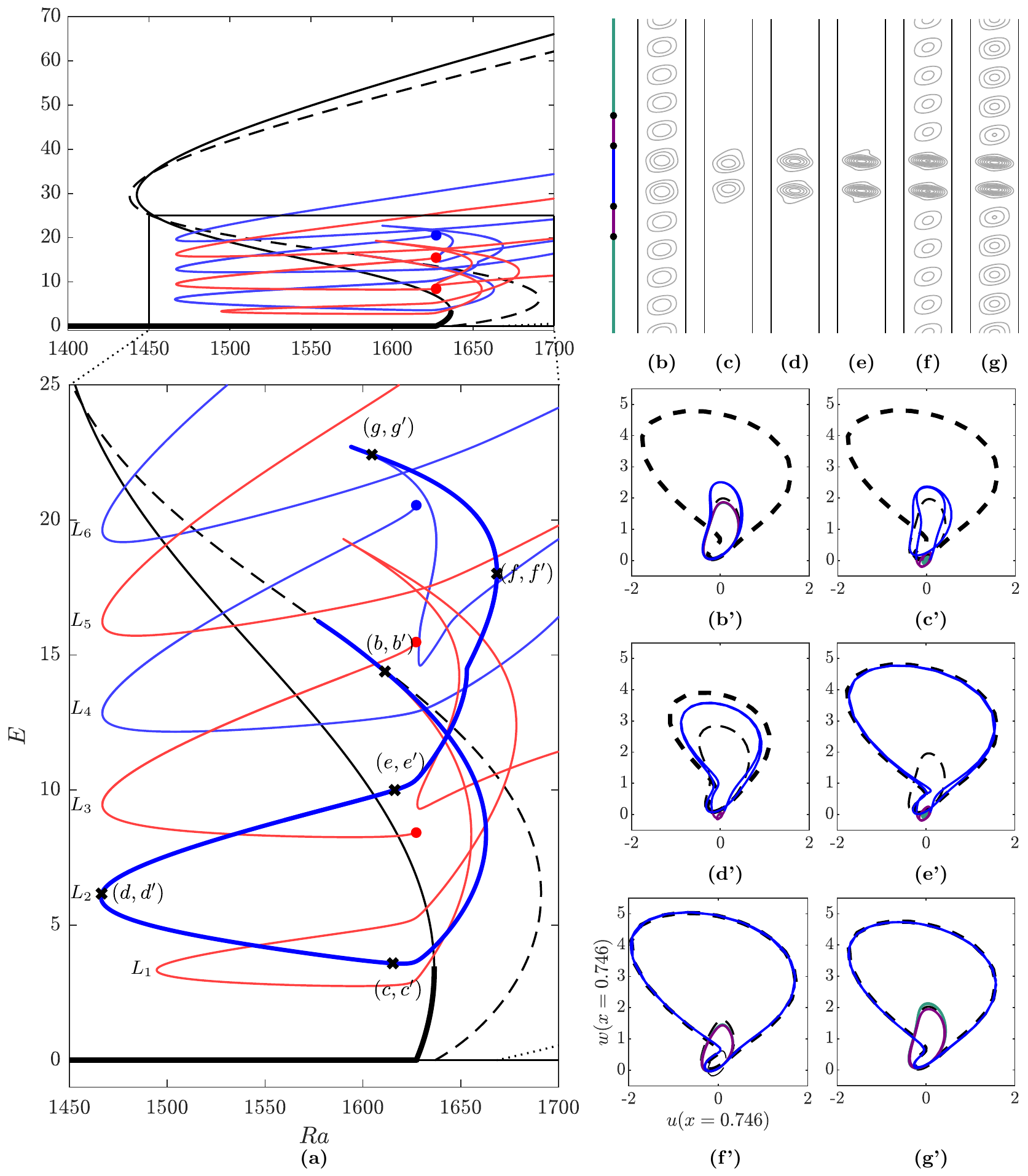}
	\caption[Bifurcation diagram, streamfunction profiles and phase-space representations for ${Pr = 0.06}$]{(a) Bifurcation diagrams for ${Pr = 0.06}$ (Stage~5), showing: the conduction state, P12 (black solid), P11 (black dashed) and $L_i$, branch of convectons with $i$ rolls (blue (red) when $i$ is even (odd)).
		The branches $L_i$ have been terminated either at ${Ra = 1700}$, or at the marked points, for clarity.
		(b)\textendash{}(g) Streamfunction profiles of the labelled states of $L_2$ and (b')\textendash{}(g') corresponding phase-space representations of these states.
		The phase-space plots show orbits representing the convectons on $L_2$ (blue), states on the upper (thick black dashed) and middle (thin black dashed) branch segments of P11 at the same value of the Rayleigh number.}
	\label{fig:5_bdlz12le5pr006}
\end{figure}

We now consider the supercritical regime:\linebreak ${Pr < Pr_c\approx 0.062}$. 
The primary branch no longer undergoes an Eckhaus instability shortly after onset, as we showed in section~\ref{sec:5_secbif}, and we cease finding the small-amplitude snaking seen in figure~\ref{fig:5_P12secbif} that was associated with convectons containing R1 rolls.
However, the large-amplitude convectons with R2 rolls that developed in Stage 2 persist into this regime and continue to lie on an organised set of disconnected branches ($L_i$ for $i=1,...,6$) that exists over both a range of both sub- and supercritical Rayleigh numbers, as evidenced when ${Pr = 0.06}$ in figure~\ref{fig:5_bdlz12le5pr006}(a).

Figures~\ref{fig:5_bdlz12le5pr006}(b,b')\textendash{}(g,g') indicate the changes that the steady states on $L_2$ undergo as the branch is followed away from its bifurcation point on P11 at ${Ra \approx 1577}$, along with $L_1$.
States on $L_2$ initially contain eleven rolls, whose form resemble rolls on states on the middle segment of P11.
As the branch is followed away from the secondary bifurcation towards larger Rayleigh numbers, the two central rolls strengthen, whilst the nine background rolls weaken in line with the middle P11 state to give states like the one shown in (b) and (b').
The background rolls continue to weaken as the branch turns around to lower Rayleigh numbers at ${Ra \approx 1663}$, so that when $L_2$ re-enters the subcritical region ${Ra < Ra_c}$, around the point marked (c), convectons contain two stronger rolls within a nearly quiescent background.
These stronger rolls (blue trajectories in the phase-space plots) resemble those on the middle portion of P11 (thin black dashed trajectory), albeit with larger amplitude, as seen in (c').
These central rolls strengthen to resemble rolls in states on the upper P11 branch (thick black dashed trajectory) as $L_2$ is followed towards the left saddle node at ${Ra \approx 1467}$ (d,d') and back to ${Ra \approx Ra_c}$ on the following upper branch segment (e,e').
As the branch traverses the critical Rayleigh number and heads towards the saddle node marked (f), nine weak rolls strengthen uniformly outside the two central rolls, thereby replacing the now-unstable conduction state, a behaviour that was observed in the cubic-quintic-septic Swift\textendash{}Hohenberg equation \cite{knobloch2019defectlike} and in Faraday experiments \cite{arbell2000}.
The background rolls increase in amplitude as the Rayleigh number increases and ultimately resemble rolls in states on the unstable middle segment of the P11 branch, which can also be seen by the purple and green trajectories in the phase-space plots (f') and (g'), which represent these background rolls, following the thin black dashed trajectory, which represent states on the P11 branch.
The branch $L_2$ continues after the saddle node near (g); however, the subsequent behaviour is not considered here as the background rolls become non-uniform and the states become two-pulsed.

States on $L_1$ display analogous behaviour across the branch segment shown in figure~\ref{fig:5_bdlz12le5pr006}(a), except with a single central roll instead of a pair of rolls.
In contrast, on the other convecton branches ($L_i$ for $i = 3,...,6$), only the outer two rolls adjust their shape along segments analogous to (c,c')\textendash{}(e,e') on $L_2$, while the interior rolls continue to follow the form of states on the upper P11 branch.
The background conduction state continues to fill with rolls as the branches cross into supercritical Rayleigh numbers (${Ra > Ra_c}$), which may be deduced from figure~\ref{fig:5_bdlz12le5pr006}(a) as the branches change gradient when they cross ${Ra = Ra_c}$.

\subsection{Low Prandtl number convectons}
There remains the outstanding question as to what happens to these convectons and the branches on which they lie as the Prandtl number decreases beyond ${Pr = 0.06}$.
We were unable to address this here owing to the numerical difficulties associated with thin viscous boundary layers at small Prandtl numbers, but we can briefly speculate.
The primary bifurcation will become increasingly supercritical and we anticipate that the large-amplitude subcritical saddle nodes of P11 and P12 will move towards higher Rayleigh numbers and, ultimately, beyond the primary bifurcation at ${Ra=Ra_c}$.
At this point, the primary branches will lie entirely within the supercritical regime ${Ra > Ra_c}$, as we found for parameter values in \cite{beaume2022near} (see Region 4 parameter values).
We suspect that the organised structure of the disconnected convecton branches will persist and shift towards larger Rayleigh numbers and we believe that these branches will continue to lie to the right of the large-amplitude saddle nodes of the primary branches.
As these convecton branches would lie within the supercritical regime, where the conduction state is unstable, we might expect that the corresponding localised states consist of a number of strong central R2 rolls within a background of weaker rolls, similarly to figures~\ref{fig:5_bdlz12le5pr006}(f) and (g).

\section{Summary and discussion}
In this paper, we investigated spatially localised steady states in natural doubly diffusive convection and, more particularly, we characterised how the structure of the branches that carry them change when the Prandtl number is varied.
We found that over a range of moderate values of the Prandtl number these branches are intertwined in parameter space in a behaviour referred to as homoclinic snaking.
This snaking behaviour was found to increase in complexity at lower Prandtl numbers, which we mostly attributed to the transition between convectons containing buoyancy-driven rolls (referred to as R1 rolls) and those containing rolls driven by a balance between inertia and buoyancy (referred to as R2 rolls).
As spatially periodic states with both types of rolls coexist with the stable conduction state over a range of Prandtl numbers, we anticipate that our results will be of interest to the study of other systems that exhibit tristability \cite{gandhi2018spatially,knobloch2019defectlike}.
We further showed that natural doubly diffusive convection provides an example of a non-variational system without a conserved quantity that admits localised states when the primary bifurcation is supercritical. which is a property that has only recently been observed in the simpler variational cubic-quintic-septic Swift\textendash{}Hohenberg equation \cite{knobloch2019defectlike}.
%

Our work is part of a ongoing effort to understand spatially localised patterns in more realistic convection systems \cite{beaume2020transition,knobloch2015,jacono2017localized,mercader2019,watanabe2016}.
The presence of symmetries in our system is obviously a feature we do not expect to hold in experiments or in a natural setting, so a natural perspective to this work is to consider such effects.
For example, the effect of boundary condition asymmetry was reported in \cite{jacono2017localized}, while \cite{mercader2019} investigated the influence of domain inclinations on convectons in doubly diffusive convection driven by vertical gradients of buoyancy.
We are currently investigating symmetry-breaking effects generated by a general choice of buoyancy ratio $N \ne -1$ \cite{tumelty2022localised} and will report on that elsewhere.

The stability of the primary and secondary branches is also of interest when considering convection patterns in physical systems.
While we have not considered such an analysis in detail here, we should note that each primary branch destabilises in a drift-pitchfork bifurcation \cite{beaume2022near} and that this occurs on the lower segments of the primary branches when ${Pr < 0.11}$.
As a result, convectons in low Prandtl number regimes can exist despite the lack of bistability between the conduction state and large-amplitude periodic states.
The stability analysis of primary branches also raises an interesting question about whether convectons inherit the drift instability. 
If such instabilities occur, it could be of interest to investigate the associated properties since symmetry-breaking instabilities of three-dimensional convectons can lead to chaotic behaviour \cite{beaume2020transition,beaume2018three}.

\section*{Acknowledgements}
This work was undertaken on ARC3, part of the High Performance Computing facilities at the University of Leeds, UK.

\bibliographystyle{siamplain}
\bibliography{references}

\end{document}